\pdfoutput=1

\newcommand*{\ATLASLATEXPATH}{latex/}

\documentclass[cernpreprint,texlive=2016,txfonts,UKenglish]{\ATLASLATEXPATH atlasdoc}

\usepackage[subfigure]{\ATLASLATEXPATH atlaspackage}

\usepackage{multirow}

\usepackage{\ATLASLATEXPATH atlasbiblatex}

\usepackage{\ATLASLATEXPATH atlasphysics}

\usepackage{tabularx}

\graphicspath{{logos/}{figures/}}

\usepackage{TIDEPaper-defs}

\AtlasTitle{Performance of the ATLAS track reconstruction algorithms in dense environments in LHC Run 2}

\PreprintIdNumber{CERN-EP-2017-045}

\AtlasJournal{EPJC}
\AtlasJournalRef{\EPJC\ 77 (2017) 673}
\AtlasDOI{10.1140/epjc/s10052-017-5225-7}

\AtlasAbstract{%
With the increase in energy of the Large Hadron Collider to a centre-of-mass energy of 13~\TeV\ for Run 2, events with dense environments, such as in the cores of high-energy jets, became a focus for new physics searches as well as measurements of the Standard Model. These environments are characterized by charged-particle separations of the order of the tracking detectors sensor granularity. 
Basic track quantities are compared between 3.2~fb$^{-1}$ of data collected by the ATLAS experiment and simulation of proton--proton collisions producing high-transverse-momentum jets at a centre-of-mass energy of 13~\TeV. 
The impact of charged-particle separations and multiplicities on the track reconstruction performance is discussed. 
The track reconstruction efficiency in the cores of jets with transverse momenta between 200 \GeV\ and 1600 \GeV\ is quantified using a novel, data-driven, method.
The method uses the energy loss, \ensuremath{\textrm{ d\textit{E}/d\textit{x}}}, to identify pixel clusters originating from two charged particles.
Of the charged particles creating these clusters, the measured fraction that fail to be reconstructed is $0.061 \pm 0.006\ \textrm{(stat.)} \pm 0.014\ \textrm{(syst.)}$ and $0.093 \pm 0.017\ \textrm{(stat.)}\pm 0.021\ \textrm{(syst.)}$ for jet transverse momenta of 200--400 \GeV\ and 1400--1600 \GeV, respectively. 
}

\hypersetup{pdftitle={Performance of the ATLAS track reconstruction algorithms in dense environments in LHC Run 2},pdfauthor={The ATLAS Collaboration}}

\begin{document}

\maketitle

\newcommand{\AtlasCoordFootnote}{%
ATLAS uses a right-handed coordinate system with its origin at the nominal interaction point (IP)
in the centre of the detector and the $z$-axis along the beam pipe.
The $x$-axis points from the IP to the centre of the LHC ring,
and the $y$-axis points upwards.
Cylindrical coordinates $(r,\phi)$ are used in the transverse plane, 
$\phi$ being the azimuthal angle around the $z$-axis.
The pseudorapidty is defined in terms of the polar angle $\theta$ as $\eta = -\ln \tan(\theta/2)$.
Angular distance is defined as $\Delta R \equiv \sqrt{(\Delta\eta)^{2} + (\Delta\phi)^{2}}$.}

\section{Introduction}
\label{sec:intro}
The Large Hadron Collider (LHC) entered a new energy regime in 2015, at the start of Run 2, with proton--proton collisions at a centre-of-mass energy of 13~\TeV. Events with \TeV-scale jets 
showering in the detectors, or $\tau$-leptons and $b$-hadrons that pass through 
multiple active layers of material, now occur at high enough rates to be studied in detail. 
These signatures also occur in potential new physics scenarios including massive new resonances decaying to highly boosted bosons or top quarks whose decay products are often reconstructed as a single jet~\cite{Butterworth}.
In the cores of highly energetic hadronic jets and $\tau$-leptons, 
the average separation between highly collimated charged particles is comparable 
to the granularity of individual sensors of the inner detector. This can create confusion within the algorithms 
used to reconstruct charged-particle trajectories, or \textit{tracks}. Therefore, without careful consideration, the 
track reconstruction efficiency in these dense environments is limited, resulting 
in difficulties in identifying long-lived \textit{b}-hadrons and hadronic $\tau$-decays, or in 
calibrating the energy and mass of jets. To prevent losses in efficiency, to increase
the possibility of discovering new phenomena and to allow more detailed
measurements of the newly opened kinematic regime, a dedicated optimization for dense 
environments was performed and deployed in the ATLAS~\cite{PERF-2007-01} reconstruction 
for the start of Run 2. This updated reconstruction provides superior physics performance, reduces the required computing resources, and is now the default used by ATLAS.

This paper first describes the ATLAS detector (Section~\ref{sec:detector}).
Then, a general overview of the track reconstruction algorithm (Section~\ref{sec:tracking})
is given, focusing on the performance of charged-particle reconstruction in dense 
environments at the start of Run 2.  The data set utilized is described in Section~\ref{sec:samples}.
The quality of the expected performance is evaluated in dedicated 
single-particle and dijet simulation samples (Section~\ref{sec:expectedPerformance}),
and comparisons between simulation and data are performed
in events with energetic jets. Extending these mainly Monte Carlo (MC) simulation-based studies, a fully data-driven method is introduced in Section~\ref{sec:dEdx} which probes the fraction of tracks lost in reconstruction, due to the high density and collimation of charged particles in high-transverse-momentum\footnote{\AtlasCoordFootnote} (\pT) jets. This is achieved by using the ionization energy loss (\dEdx) in the pixel detector.

\section{The ATLAS detector}
\label{sec:detector}
The ATLAS experiment, a multipurpose particle detector at the LHC, covers almost the entire solid angle around the collision point, and consists of an inner detector (ID) tracking system surrounded by a thin superconducting solenoid magnet producing a 2 T axial magnetic field, electromagnetic and hadronic calorimeters, and a muon spectrometer incorporating three large toroid magnet assemblies.

\begin{figure}[htbp]
\centering
\includegraphics[width=0.65\textwidth]{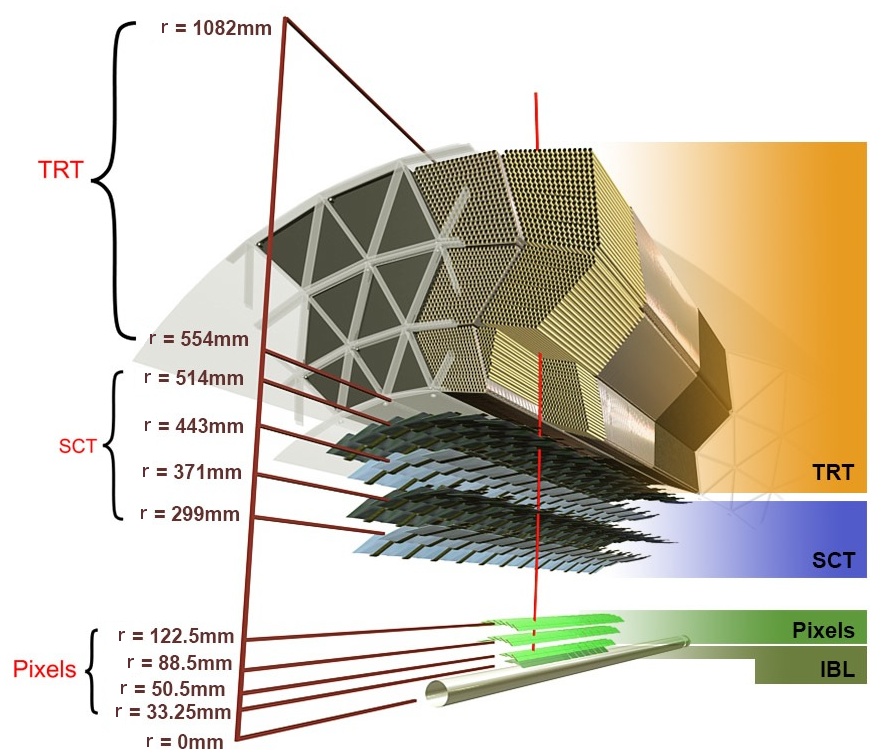}\textsl{\textsl{}}
\caption[Sketch of the barrel region of the ATLAS inner detector.]{\label{fig:IDSketch} 
}
Sketch of the barrel region of the ATLAS inner detector.
\end{figure}

The ID, shown in Figure~\ref{fig:IDSketch}, provides position measurements for charged particles in the 
range $|\eta| < 2.5$ by combining information from three subdetectors. It consists of a cylindrical barrel region (full coverage for $|\eta| \lesssim 1.5$) arranged around the beam pipe, and two end-caps. Disks in the end-cap region are placed perpendicular to the beam axis, covering $1.5 < |\eta| < 2.5$.
Starting from the interaction point,
the high-granularity silicon pixel detector segmented in $r$--$\phi$ and $z$ (including the 
new innermost layer, the insertable B-layer (IBL)~\cite{ATLAS-TDR-19} added for Run 2) 
covers the vertex region and 
typically provides four measurements per track.
The IBL has a mean radius of 33~mm and
a typical IBL pixel has a size of 50~$\mu$m by 250~$\mu$m in 
the transverse  and longitudinal directions with a sensor thickness of 200~$\mu$m.
For the remaining three layers of the pixel system, located
at mean radii of 50.5, 88.5, and 122.5~mm respectively,
a typical pixel has a size of 50~$\mu$m by 400~$\mu$m in the transverse and longitudinal directions
with a thickness of 250~$\mu$m. The pixel layer at a radius of 50.5~mm is referred to as the B-layer in this paper. The coverage in the end-cap region is enhanced by three disks on either side of the interaction point.
The pixel detectors measure the charge collected in each individual pixel using the time over threshold~(ToT)~\cite{Aad:1119279}. ToT is the time the pulse exceeds a given threshold and is proportional to the deposited energy.

Outside the pixel volume, the barrel of the silicon microstrip detector (SCT) consists of 
four double strip layers at radii of 299 mm to 514 mm, complemented by nine disks in each of the end-caps. A typical strip of a barrel SCT sensor has a length of 126 mm and a pitch of 80~$\mu$m.
On each layer, the strips are parallel to the beam direction on one side 
and at a stereo angle of 40~mrad on the other.
The information from the two sides of each layer can be combined to provide an average of four three-dimensional measurements per track.
The SCT sensors are connected to binary read-out chips, which do not provide
information about the collected charge.
The silicon detectors are complemented by the transition radiation tracker (TRT)~\cite{1748-0221-3-02-P02014},
which extends track reconstruction radially up to a radius of 1082 mm for 
charged particles within $|\eta| = 2.0$
while providing $r$--$\phi$ information. The raw timing information
from its straw tubes
is translated into calibrated drift circles that are matched to track
candidates reconstructed from the silicon detectors~\cite{1748-0221-3-02-P02014}.

The solenoid is surrounded by sampling calorimeters.
Calorimetry is provided by three distinct detectors outside the ID volume.
A lead/liquid-argon sampling electromagnetic calorimeter is split into barrel ($|\eta|$ < 1.5) and end-cap ($1.5< |\eta| < 3.2$) sections.  A steel/scintillator-tile hadronic calorimeter covers the barrel region ($|\eta| < 1.7$) and two end-cap copper/liquid-argon sections extend to higher pseudorapidity ($1.5 < |\eta| < 3.2$).  Finally, the forward region ($3.1 < |\eta| < 4.9$) is covered by a liquid-argon calorimeter with a copper (tungsten) absorber in the electromagnetic (hadronic) section.  
In the outermost part, air-core toroids provide the magnetic field for the muon spectrometer. It consists of three layers of gaseous detectors: monitored drift tubes and cathode strip chambers for muon identification and momentum measurements for $|\eta| < 2.7$, and resistive-plate and thin-gap chambers for online event selection up to $|\eta| = 2.4$. A two-level trigger system, custom hardware followed by a software-based level, is used for online event selection and to reduce the event rate to about 1 kHz for offline reconstruction and storage.

\section{ATLAS track reconstruction}
\label{sec:tracking}

The following 
provides an overview of primary-track reconstruction in the pixel and SCT detectors.
After cluster creation, the primary-track reconstruction algorithm 
utilizes iterative track-finding seeded from combinations of silicon detector measurements,
while additional methods are employed to recover non-prompt tracks. 
A staged pattern-recognition approach is used: a loose track candidate search, which allows a number of combinatorial track candidates, is followed by a stringent ambiguity-solver that compares and rates the individual tracks by assigning a relative \textit{track score} to each track. This follows current approaches to track reconstruction first introduced 
in Ref.~\cite{Wicke}. 
Further details, including a description of TRT track extensions, can be found in Ref.~\cite{Cornelissen:2008zzc}.

\subsection{Clusterization}
\label{sec:clusteriation}
Charged-particle reconstruction in the pixel and SCT detectors begins
by assembling clusters from the raw measurements. A connected component analysis (CCA)~\cite{atlasCCA} groups pixels and strips in a given sensor, where the deposited energy yields a charge
above threshold, with a common edge or corner into clusters. 
From these clusters, three-dimensional measurements referred to as \textit{space-points} are created. 
They represent the point where the charged particle traversed the active material of the ID.
In the pixel detector, each cluster equates to one space-point, while
in the SCT, clusters from both sides of a strip layer must be combined
to obtain a three-dimensional measurement.

The charge in a pixel sensor is often collected on multiple adjacent pixels. In the data set described in Section~\ref{sec:samples}, the average size of pixel clusters in the barrel is about two pixels in the $r-\phi$ plane and from one to three pixels in the longitudinal direction increasing with $\eta$. 
The total charge is proportional to the path length in the sensor
and thus dependent on the incident angle of the particle.
The particle's intersection point with the sensor 
is determined from the pixels contributing to the cluster using a linear approximation refined
with a charge interpolation technique~\cite{BELAU1983253}.
In dense environments, the spatial separation between charged particles traversing the sensor 
is only a few pixels, and the CCA algorithm, at times, reconstructs only one cluster
which includes energy deposits from multiple particles. Identifying such clusters reliably and quickly is paramount for an efficient charged-particle reconstruction 
in dense environments.

\begin{figure}[htbp]
\centering
\subfigure[Single-particle pixel clusters] {
  \includegraphics[width=0.45\textwidth]{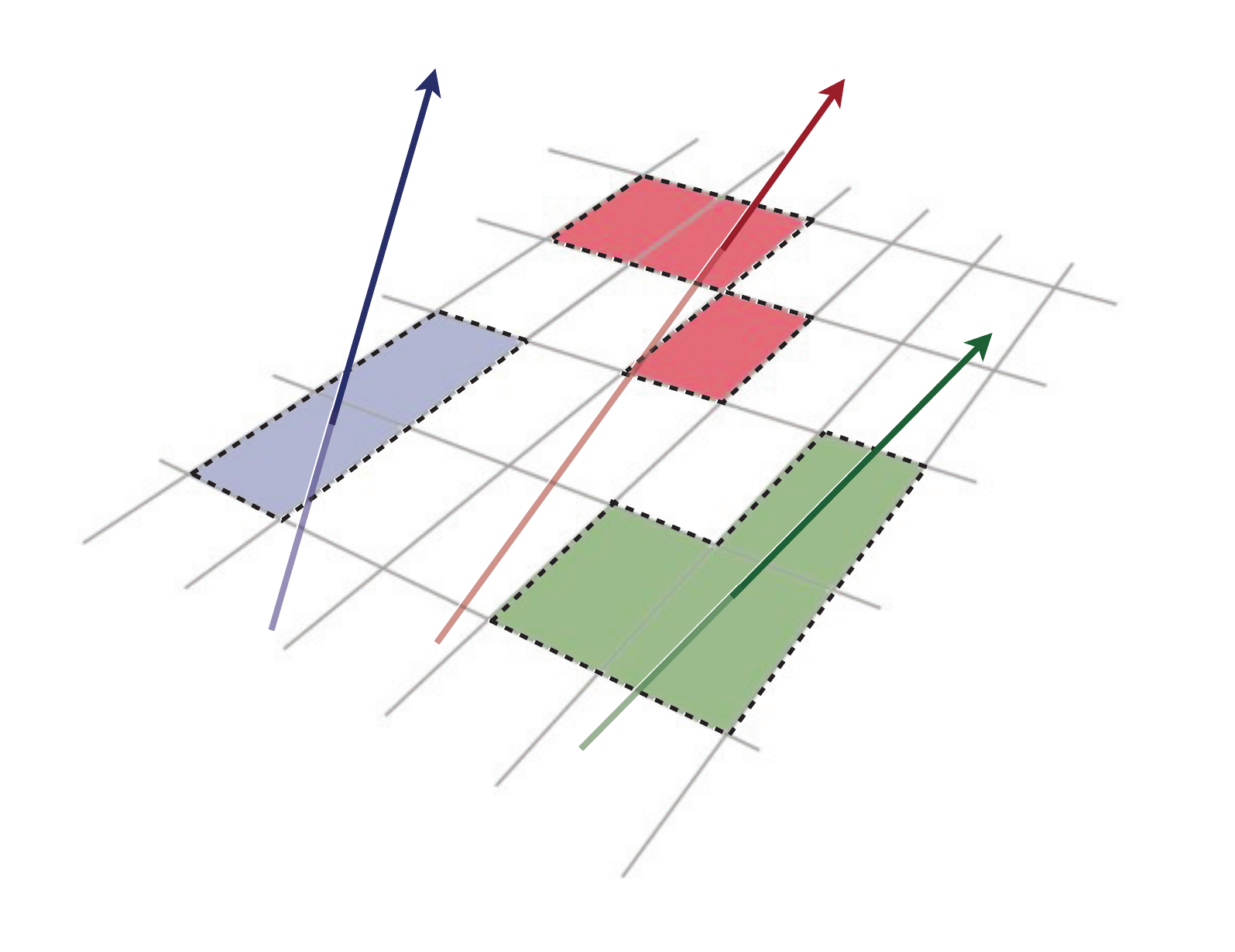}
}
\subfigure[Merged pixel cluster] {
  \includegraphics[width=0.45\textwidth]{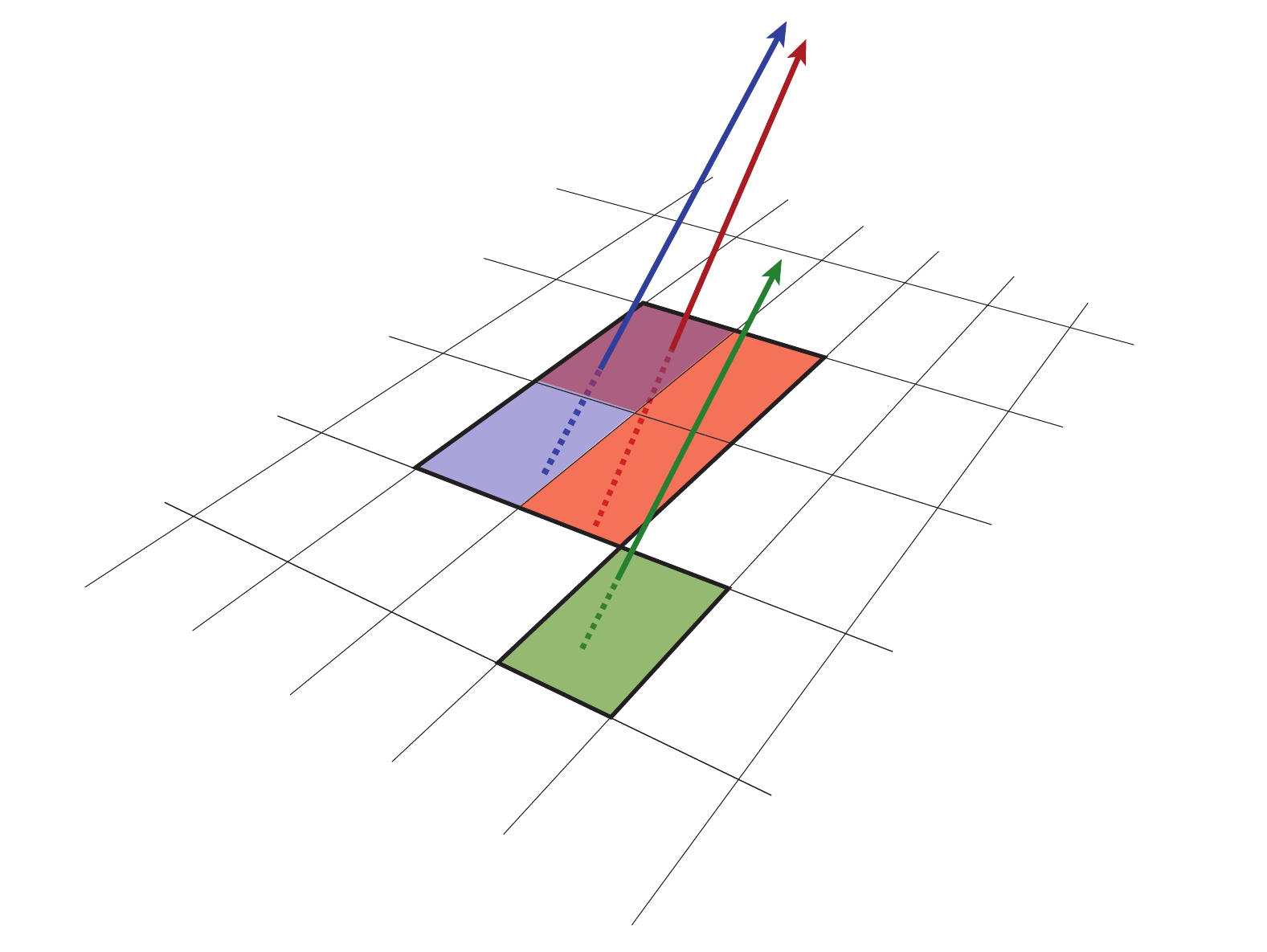}
}
\caption[Illustration of single-particle pixel clusters on a pixel module and a merged pixel clusters due to very collimated charged particles.]
{\label{fig:sketchMergedClusters}
Illustration of (a) single-particle pixel clusters on a pixel sensor and (b) a merged pixel cluster due to very collimated charged particles. Different colours represent energy deposits from different charged particles traversing the sensor and the particles trajectories are shown as 
arrows.
}
\end{figure}

It is useful to introduce the several classes of clusters 
identified by either the "truth information", only available in simulation and referring to information at MC generator level,
or reconstructed quantities in both collision data and MC simulation.
Clusters created by charge deposits from one particle are called \textit{single-particle} clusters.
Clusters created by charge deposits from multiple particles are called \textit{merged} clusters. 
These definitions rely on truth information
and both cases are illustrated in Figure~\ref{fig:sketchMergedClusters}.
Based on information available in the track reconstruction 
algorithm described below, clusters which are compatible with a merged cluster
can be identified. These are labelled \textit{identified as merged}. Ideally,
all clusters identified as merged are, in fact, merged clusters, and all merged clusters are 
identified as merged.
\textit{Shared} clusters are those which are used in multiple reconstructed
tracks but are not sufficiently compatible with the properties of a 
merged cluster to be identified as merged by the reconstruction.
\textit{Multiply used} clusters - clusters used by multiple tracks - are either identified as merged or shared but not both.

\subsection{Iterative combinatorial track finding}
\label{sec:seedfind}
Track seeds are formed from sets of three space-points. This approach maximizes the possible number of combinations while still allowing a first crude momentum estimate.  The impact parameters of a track seed, with respect to the centre of the interaction region, are estimated by assuming a perfect helical trajectory in a uniform magnetic field. 

The purity, or fraction of seeds that result in good-quality tracks, 
varies significantly depending on which subdetector(s) recorded the space-points used in the seed. Therefore, seed types are considered starting with SCT-only, then pixel-only and finally mixed-detector seeds, representing the order of purity.
A number of criteria are placed on the seeds to maximize purity: first and foremost seed-type-dependent momentum and impact parameter requirements.  
Also, the use of space-points in multiple seeds is carefully controlled.
Purity is further improved by requiring that one additional space-point is compatible with the particle's trajectory estimated from the seed.
A combinatorial Kalman filter~\cite{Rudi} is then used to build track candidates from the 
chosen seeds by incorporating additional space-points from the remaining layers of the pixel and SCT detectors which are compatible with the preliminary trajectory.
The filter creates multiple track candidates per seed if more than one
compatible space-point extension exists on the same layer.

These criteria result in a very high efficiency for reconstructing primary particles (for example, the muon reconstruction efficiency is greater than 99\%~\cite{PERF-2015-10}) and 
the removal of tracks created from purely random 
collections of space-points.
Suppressing such purely combinatorial tracks is essential in order to remain within the available CPU budget for event reconstruction. 
From approximately 13 space-point combinations created for an isolated charged particle traversing the entire ID, the time-intensive combinatorial Kalman filter is, on average, called in its entirety 1.1 times.
As all realistic combinations of space-points have been made, there are a number of track candidates where space-points overlap, or have been incorrectly assigned. This necessitates an ambiguity-solving stage. 

\subsection{Track candidates and ambiguity solving}
\label{sec:ambisolving}

In the ambiguity solver, track candidates 
considered to create the reconstructed track collection
are processed individually
in descending order of a track score, favouring tracks with a higher score. 
This design relies on having an appropriate track score definition
that puts tracks into an order that scores more highly the candidates likely to correctly represent the trajectory of a charged primary particle.

The method used to determine the \textit{track score}, discussed in the following, 
applies a robust approach based largely on simple measures of the track quality.
Clusters assigned to a track increase the track score according to 
configurable weight fractions reflecting the 
intrinsic resolutions and expected cluster multiplicities 
in the different subdetectors.
Holes\footnote{Holes are defined as intersections of the reconstructed track trajectory with a sensitive detector element that does not contain a matching cluster.  These are estimated by following closely the track trajectory and comparing, within the uncertainties, the intersected sensors with the clusters on the track.  
Inactive sensors or regions, such as edge areas on the silicon sensors, are excluded from the hole definition.} 
reduce the score.
The $\chi^{2}$ of the track fit is also considered to penalize candidates with a poor fit.  
Finally, the logarithm of the track momentum is considered to promote energetic tracks 
and suppress the larger number of tracks with incorrectly assigned clusters, which typically have a low \pT.

After the track scores have been calculated, the ambiguity solver deals with clusters assigned to multiple track candidates.
Clusters compatible with multiple track candidates are a natural consequence of having merged
clusters in dense environments. High reconstruction efficiency is facilitated by the identification of merged clusters, as explained in Section~\ref{subsec:pixelclustering}.
However, shared clusters, clusters used in multiple track candidates which are not 
identified as merged, must be limited as they are a strong indicator of incorrect assignments. 

\begin{figure}[htbp]
\centering
\includegraphics[width=0.95\textwidth]{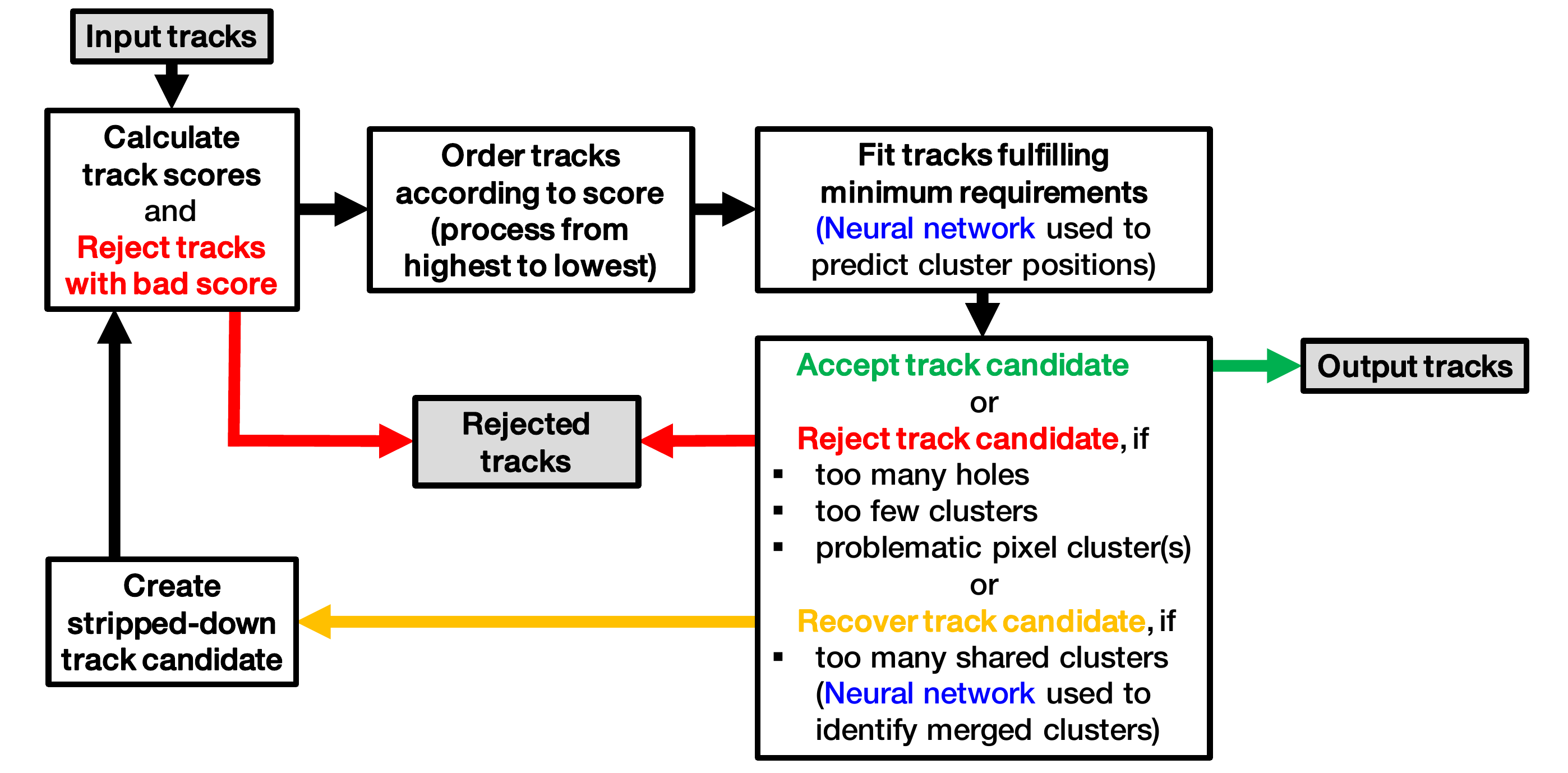}\textsl{\textsl{}}
\caption[Flow of track candidates through the ambiguity solver.]{\label{fig:ambiSketch} 
}
Sketch of the flow of tracks through the ambiguity solver.
\end{figure}

To count shared clusters, a track candidate is only compared to those
tracks previously accepted by the ambiguity solver. 
Clusters can be shared by no more than two 
tracks, giving preference to tracks processed first in the ambiguity solver.
Also, a track can have no more than two shared clusters.  
A cluster is removed from a track candidate if it causes either the candidate or an accepted track to not meet the shared-cluster criterion.
The track candidate is then scored again and returned to the ordered list of remaining candidates.
Track candidates are rejected by the ambiguity solver 
if they fail to meet any of the following basic quality criteria:
\begin{itemize}
\item $\pT > 400$ \MeV,
\item $|\eta| < 2.5$,
\item Minimum of 7 pixel and SCT clusters (12 are expected),
\item Maximum of either one shared pixel cluster or two shared SCT clusters on the same layer,
\item Not more than two holes in the combined pixel and SCT detectors,
\item Not more than one hole in the pixel detector,
\item $|d_0^{\textrm{BL}}| <$ 2.0 mm,
\item $|z_0^{\textrm{BL}}\sin\theta| <$ 3.0 mm,
\end{itemize}
where $d_0^{\textrm{BL}}$ is the transverse impact parameter calculated with respect to the measured beam-line position, $z_0^{\textrm{BL}}$ is the longitudinal difference along the beam line between the point where $d_0^{\textrm{BL}}$ is measured and the primary vertex\footnote{All events considered in this analysis are required to have at least one reconstructed primary vertex with at least two associated tracks~\cite{atlas-vertex}. Only tracks compatible with the primary vertex having the highest sum of the squared transverse momenta of its associated tracks are considered.}, and $\theta$ is the polar angle of the track. In the remainder of the paper, all studied tracks fulfil these requirement.
A simplified flow of track candidates through the ambiguity solver is shown in Figure~\ref{fig:ambiSketch}.

\subsection{Neural--network pixel clustering}
\label{subsec:pixelclustering}

To aid the ambiguity solver and minimize the loss of efficiency due to limitations on the number of shared clusters per track, an artificial neural network (NN) trained to identify merged clusters is used.
The measured charge, which is proportional to the deposited energy, and relative position of pixels in the cluster can be
used to identify merged clusters. Additional information about the particle's incident angle, provided from the track candidate, significantly improves the NN’s performance~\cite{PERF-2012-05}. 
For merged clusters created by two charged particles, the NN identification efficiency of this cluster as being created by two particles is about 90\%. Merged clusters created by three charged particles are identified as such with an efficiency of 85\%. Only a few percent of single particle clusters are incorrectly identified as a two-particle merged cluster and a negligible amount are identified as three-particle merged clusters. The NN is not able to distinguish clusters from exactly three and more than three charged particles.
It is not possible for the NN to separate the energy deposits of each charged particle in an identified merged cluster and subsequently divide it into multiple clusters. 
Unlike the Run-1 reconstruction algorithm~\cite{Cornelissen:2008zzc}, the NN is 
consulted only when a cluster is used in multiple track candidates 
largely mitigating the impact of misidentification of merged clusters by the NN.

The inherent randomness of charged-particle interactions
with thin silicon layers prevents the NN from performing perfectly. 
For example, the emission of $\delta$-rays causes difficulties as
they can lead to bigger clusters and larger energy deposits than expected from a single particle. 
These inefficiencies can be mitigated by correlating information from consecutive
layers of the pixel detector.  
In general, the separation between collimated charged particles increases as
they travel outward through the ID.  
Therefore, if a pair of tracks uses a merged cluster on a given layer,
then the inner layer is likely to contain a merged cluster as well.
Furthermore, both clusters should be used by the same track candidates in this logic.

In summary, a cluster can be identified as merged in two ways. 
Either it is used by multiple track candidates and the NN identifies it as a merged cluster, or if two track candidates compete for clusters on two consecutive layers, the cluster on the inner layer is identified as merged if the cluster on the outer layer is identified as merged.
Clusters identified as merged are used by the competing track candidates without penalty.
Clusters which are not identified as merged, shared clusters, can still be used  
in multiple tracks but with the penalty described in Section~\ref{sec:ambisolving}.

\subsection{Track fit}
\label{subsec:trackfit}

For track candidates fulfiling the requirements listed in Section~\ref{sec:ambisolving}, a high-resolution fit is 
performed using all available information. Fitted tracks which pass 
through the ambiguity solver without modification are added to the 
final track collection. 
Delaying the track fit until this stage minimizes the number of times the fitter is called,
which is advantageous as it is a relatively CPU-intensive process. 

For the high-resolution track fits, the position and uncertainty of each cluster is determined by additional NNs~\cite{PERF-2012-05}. They predict the positions where the charged particles intersected the sensor based on the same input to the NN described in Section~\ref{subsec:pixelclustering}. The predicted number of charged particles which created the cluster determines the number of particle intersections the additional NNs predict. This decreases the discrepancy between the reconstructed cluster position and the cluster's fitted track position at the detector surface, especially for merged clusters, resulting in more precise track parameters.

\FloatBarrier

\section{Data and Monte Carlo samples}
\label{sec:samples}
Data from proton--proton collisions at $\sqrt{s}=13~\TeV$, collected during 2015 and corresponding to an integrated luminosity of 3.2 \ifb, are used in this paper. Events are selected using triggers requiring a single jet above various \pt thresholds. The minimum jet trigger \pT threshold is 100~\GeV. The numbers of events selected by the triggers were reduced by a factor depending on the instantaneous luminosity and the jet \pT threshold. This suppresses the number of low-\pT jets while keeping all events with at least one jet with \pT $> 450~\GeV$. Standard ATLAS data-quality requirements are applied to all data sets, ensuring all detectors were operational.  

The data are compared to a leading-order dijet MC sample generated with \PYTHIA 8.186~\cite{bib:pythia} with the A14 tuned parameter set~\cite{ATL-PHYS-PUB-2014-021} and the NNPDF2.3LO parton distribution function (PDF) set~\cite{bib:NNPDF}. MC samples generated with \Herwigpp 2.7.1~\cite{bib:herwigpp}, and \SHERPA 2.1~\cite{bib:sherpa} are also studied. For \Herwigpp, the UEEE5 tuned parameter set is used with the CTEQ6L1 PDF set~\cite{bib:CTEQ} and for \SHERPA, parameters corresponding to the CT10 PDF set~\cite{bib:CT10} are used. The ATLAS detector response is fully simulated~\cite{SOFT-2010-01} using the \GEANT 4 framework~\cite{GEANT4}. The average number of proton--proton interactions per bunch crossing (pile-up) was approximately
15 during the 2015 data-taking period. The expected contribution from additional
proton--proton interactions is accounted for by overlaying minimum-bias events simulated
with \PYTHIA 8. The MC samples are reweighted to match the distribution of the number of interactions per bunch crossing and then reweighted to the inclusive jet-\pt spectrum observed in collision data.
In dense environments, the impact of pile-up on the track reconstruction performance is small. The change in tracking efficiency considering only one interaction per bunch crossing to an average pile-up of 40 in the dijet MC sample for jets with a \pT above 200~\GeV\ is below 0.3\%. 

In order to perform detailed simulation-based studies on event topologies with highly collimated particles, four large MC samples,
with a single particle decaying into a set of nearby charged particles, are employed. The initial particles have different lifetimes and decay multiplicities, and are generated with a uniform transverse momentum spectrum from 10~\GeV\ to 1~\TeV\ within $|\eta|$ of 1.0.  
Topologies with two highly collimated tracks are studied in a simulated $\rho\rightarrow\pi^{+}\pi^{-}$ sample. Simulated decays of a single $\tau$-lepton to three charged hadrons ($\tau^{\pm} \rightarrow \pi^{+} \pi^{-} \pi^{\pm} \nu_{\tau}$) are used to study topologies with three charged particles.
To study the performance in topologies with higher charged-particle multiplicities, two additional samples are created; a sample containing all decays of a $B^{0}$ into multiple particles and a $\tau$-lepton decaying to a final state including five charged hadrons.

\section{Track reconstruction performance in dense environments}
\label{sec:expectedPerformance}
This section first compares basic properties of tracks inside jets in data with those in simulated dijet samples (Section~\ref{sec:basicdatamc}). Using truth-based quantities, Section~\ref{subsec:singleparticles} studies single-particle decays with collimated decay products. These relatively simple topologies allow the behaviour of the track reconstruction to be studied as a function of the momentum of the initial particle, and the spatial separation between the tracks. Section~\ref{subsec:jets} presents analogous results, but derived from a dijet MC sample of high-\pT jets.

\subsection{Classification}
\label{sec:objects}
\label{sec:trkselec}

In simulation, tracks are classified using a {\em truth-matching probability}.
It is the ratio of the weighted number of
measurements originating, at least in part, from the same simulated particle, 
to the weighted number of all measurements used in a track.
A subdetector-specific weight of ten for measurements in the pixel detector, five for the SCT and one for the TRT is used. These weights reflect the average number of expected measurements in each subdetector. 
A properly reconstructed track is required to have a truth-matching probability above 0.5. 
Such a requirement is imposed for all reconstruction efficiencies presented in this paper.

\textit{Fake} tracks are those which have a truth-matching probability below 0.5. 
Due to the careful pruning of seeds, the majority of reconstructed fake tracks are from 
the misallocation of clusters from other particles to a track and not purely random combinations of clusters. The track reconstruction procedure described in Section~\ref{sec:tracking} results in a negligible number of fake tracks in dense environments. For jets with a \pT~above 200~\GeV\ in the dijet MC sample described in Section~\ref{sec:samples}, the fraction of fake tracks is below 0.5\%. From only one $pp$ interaction per bunch crossing to an average pile-up of 40, this fraction increases by about 0.5\%, still making it negligible. Consequently, fake tracks are not discussed in further detail.

Jets are reconstructed from topological clusters~\cite{PERF-2014-07} of energy deposits in the calorimeter using the \antikt\ algorithm~\cite{Cacciari2008gp} with a radius parameter $R=0.4$ and are selected requiring a minimum jet \pT\ of 200~\GeV\ and $|\etaJet|<2.5$. Jets are corrected for the effects of non-compensating response in the calorimeter and inactive material by using energy- and $\eta$-dependent calibration factors, based on MC simulation
and $pp$ collision data. Additional corrections are applied to reduce the dependence of the jet energy measurement on the longitudinal and transverse structure of the jets and also to correct for jets that are not fully contained in the calorimeter~\cite{PERF-2016-04}.

\subsection{Data and MC simulation comparison}
\label{sec:basicdatamc}
This section gives an overview of basic properties of tracks inside jets. Data and MC simulation comparisons establish fair agreement between the two.

The average number of tracks per unit of angular area versus the angular distance from the jet axis in data and MC events is compared in Figure~\ref{fig:jetpt}. The charged-particle density in jets increases linearly with the logarithm of the jet momentum, which reflects the average number of tracks inside the jet. Moreover, most tracks are located within an angular distance of 0.05 from the jet axis. Jets in data tend to have a slightly wider distribution of reconstructed charged particles than those in simulation. 

\begin{figure}[htbp]
\centering
  \includegraphics[width=0.49\textwidth]{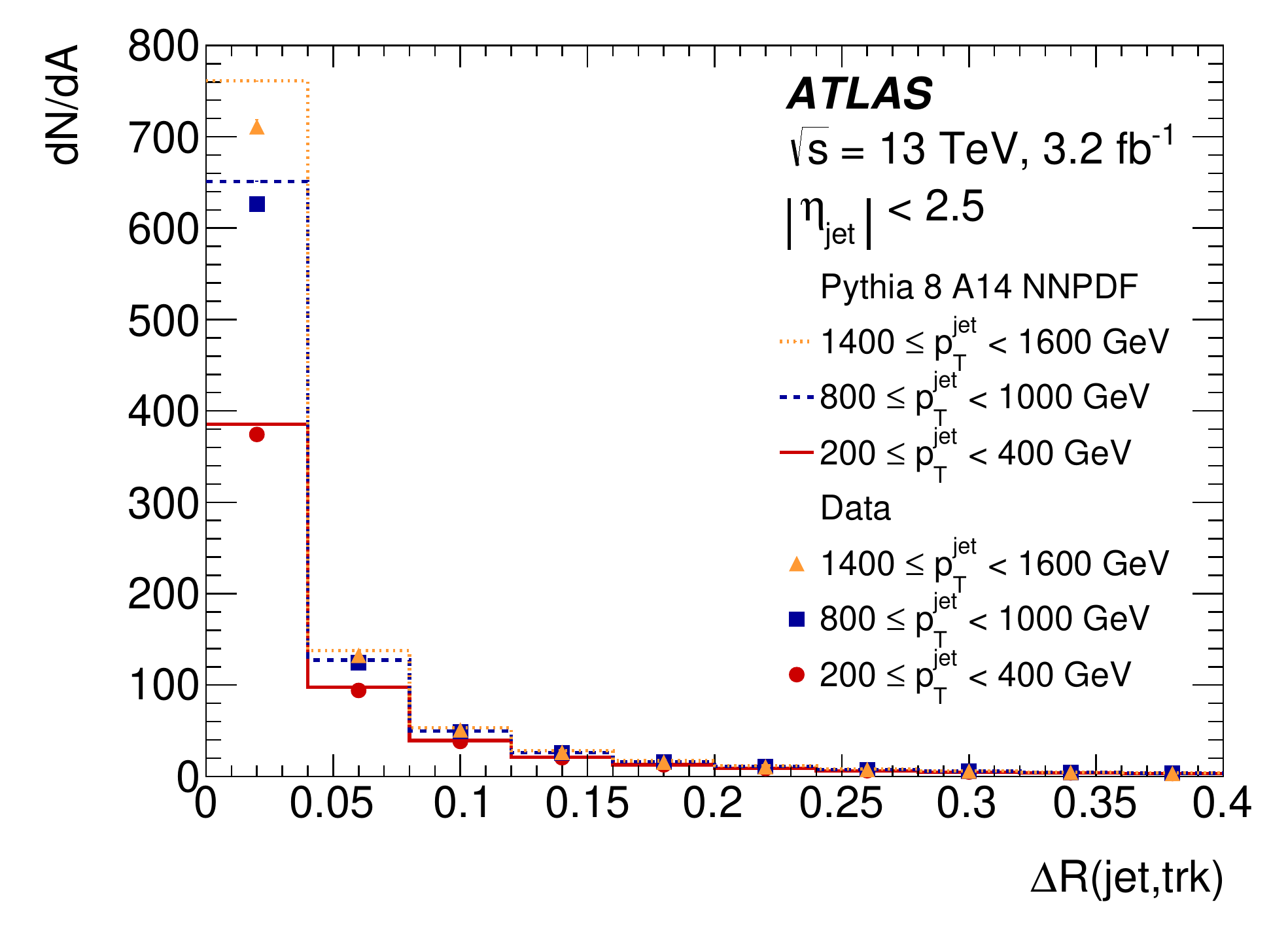}
\caption[Number of tracks versus dR(jet,trk).]{	\label{fig:jetpt} 
The average number of primary tracks per unit of angular area as a function of the angular distance from the 
jet axis. Data (markers) and dijet MC (lines) samples are compared in bins of jet \pT~showing the high density in the cores of energetic jets.}
\end{figure}

Due to the large number of collimated charged particles the number of 
\MultUsed\ clusters rises steeply at small distances to the jet axis.
Figure~\ref{fig:NPixOnTrk} shows the number of 
pixel clusters that are identified as merged and the
number of shared pixel clusters on the track 
for data and MC simulation versus the angular distance from the jet axis. 
The average number of shared pixel clusters remains relatively low compared to the number of clusters identified as merged, down to the smallest distances, because the reconstruction algorithm identifies merged 
clusters with high efficiency, and these consequently are not counted as shared.
MC simulation and data show reasonable agreement in the individual bins of jet \pT.

\begin{figure}[htbp]
\centering
\subfigure[\label{fig:NPixOnTrkb}Pixel clusters identified as merged] {\includegraphics[width=0.49\textwidth]{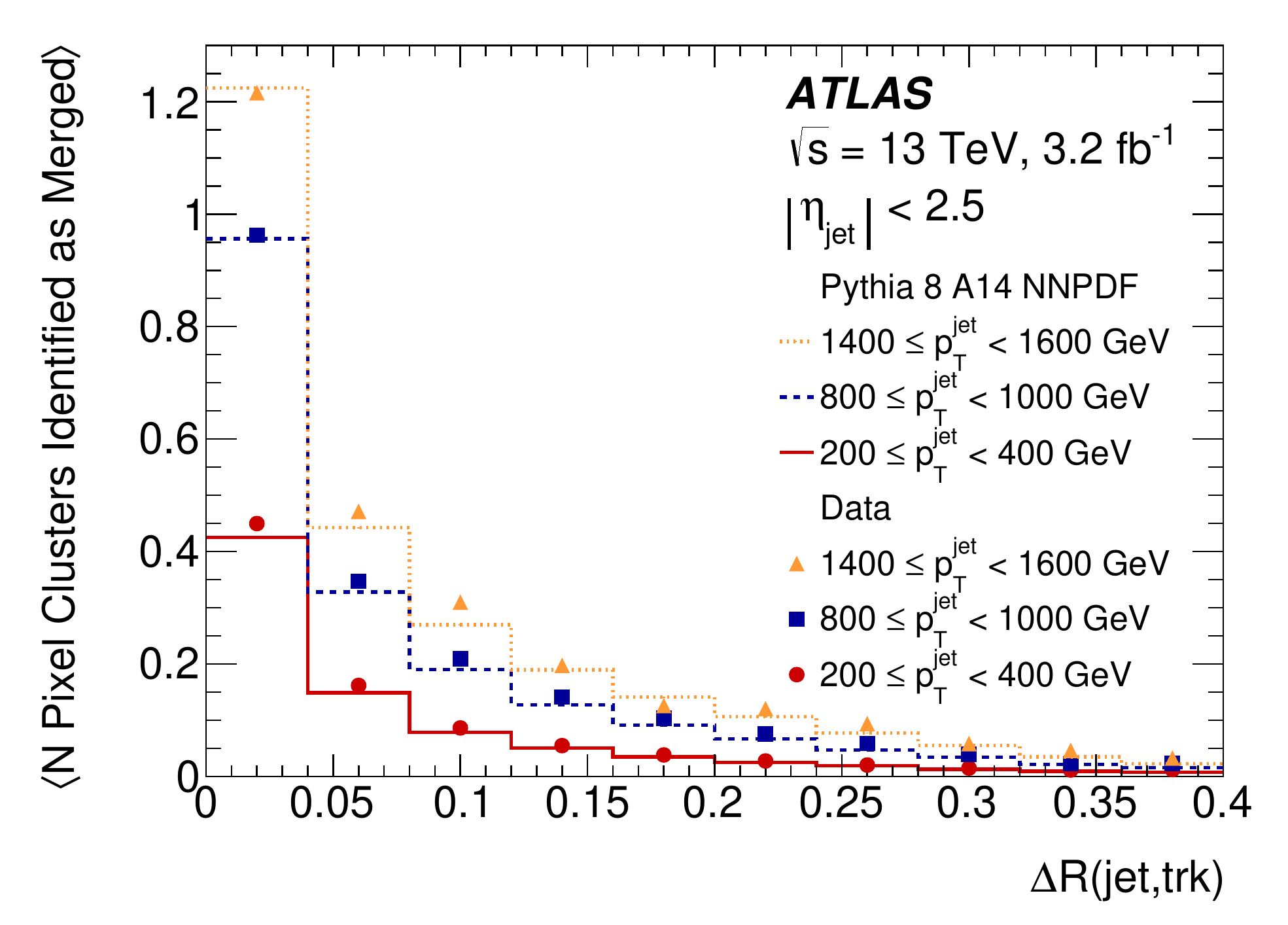}}
\subfigure[\label{fig:NPixOnTrkc}Shared pixel clusters] {\includegraphics[width=0.49\textwidth]{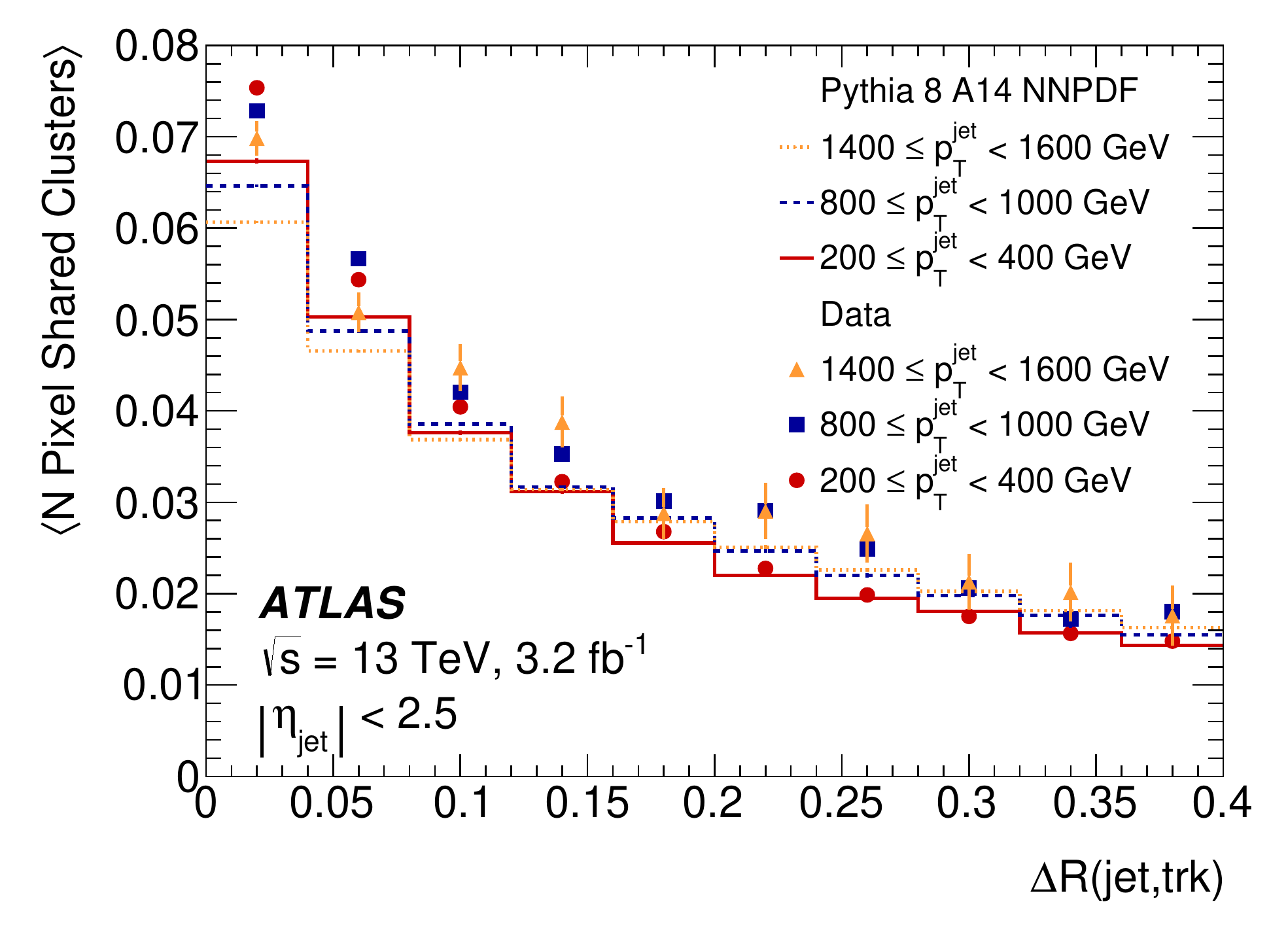}}
\caption[IBL pixel clusters.]
{\label{fig:NPixOnTrk}
The average number of (a) pixel clusters identified as merged and (b) shared pixel clusters on 
primary tracks (with a production vertex 
before the IBL) are shown as a function of the angular distance of the track from
the jet axis. Data (markers) and dijet MC (lines) samples are compared 
in bins of jet \pT.
The rise in both populations at small distances from the jet axis 
is expected due to the increasingly dense environment.
}
\end{figure}

Inefficiencies in the identification and treatment of merged clusters affect the number of IBL clusters on tracks in dense environments.
Figure~\ref{fig:NIBLOnTrk} shows the average number of IBL clusters on the track, 
for data and MC simulation versus the angular distance from the jet axis. 
For small distances the number of IBL clusters shows a drop, explained 
by a residual inefficiency in assigning clusters to the appropriate track.
MC simulation and data agree within expectations in each of the individual jet \pT\ bins. The overall lower average number of IBL clusters on track in data is due to a not fully functional IBL detector module, which is not correctly considered in MC simulation.

\begin{figure}[htbp]
\centering
\includegraphics[width=0.49\textwidth]{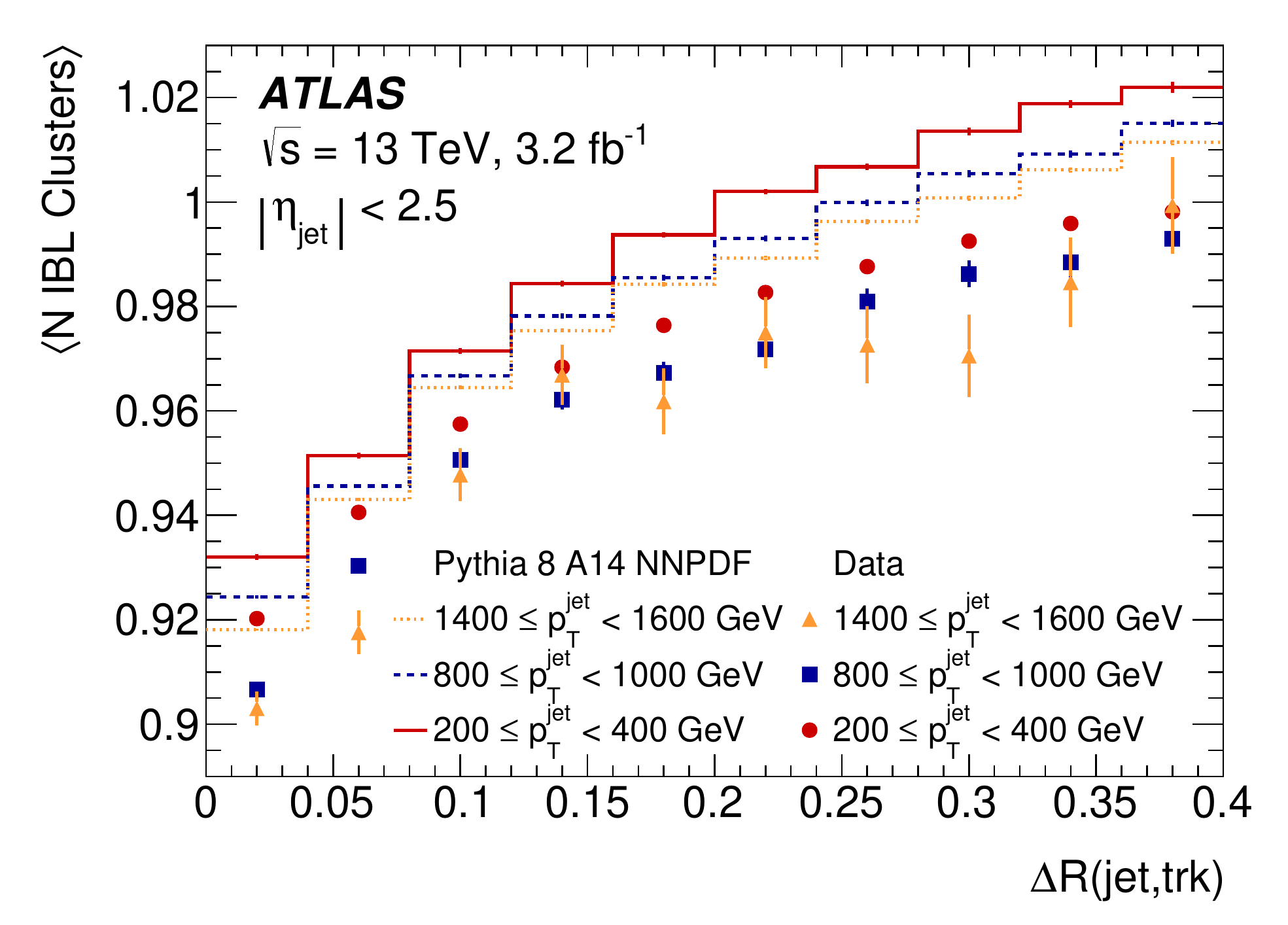}
\caption[IBL pixel clusters.]
{\label{fig:NIBLOnTrk}
The average number of IBL clusters on primary tracks (with a production vertex 
before the IBL) shown as a function of the angular distance of the track from
the jet axis. Data (markers) and dijet MC (lines)) samples are compared 
in bins of jet \pT\ showing a slight drop at small distances explained by a residual
cluster-to-track assignment inefficiency.
}
\end{figure}

Although the SCT sensors are located at much higher radii than the pixel sensors, the expected number of shared clusters is considerably larger than for the pixels as shown in Figure~\ref{fig:sctshared}. 
This is due to the coarser segmentation of the SCT strips in one dimension and the lack of charge information hindering the identification of merged SCT clusters. 
The average number of shared SCT clusters decreases with the angular distance from the jet axis, correlated with the decrease in charged-particle density visible in Figure~\ref{fig:jetpt} for data and MC simulation. 
In the studied jet-\pT range, the average number of SCT clusters on tracks 
is approximately 7.7 with little variation with respect to angular distances from the jet axis.
The MC simulation agrees within expectations with data in the individual bins of jet \pT.

\begin{figure}[htbp]
\centering
  \includegraphics[width=0.49\textwidth]{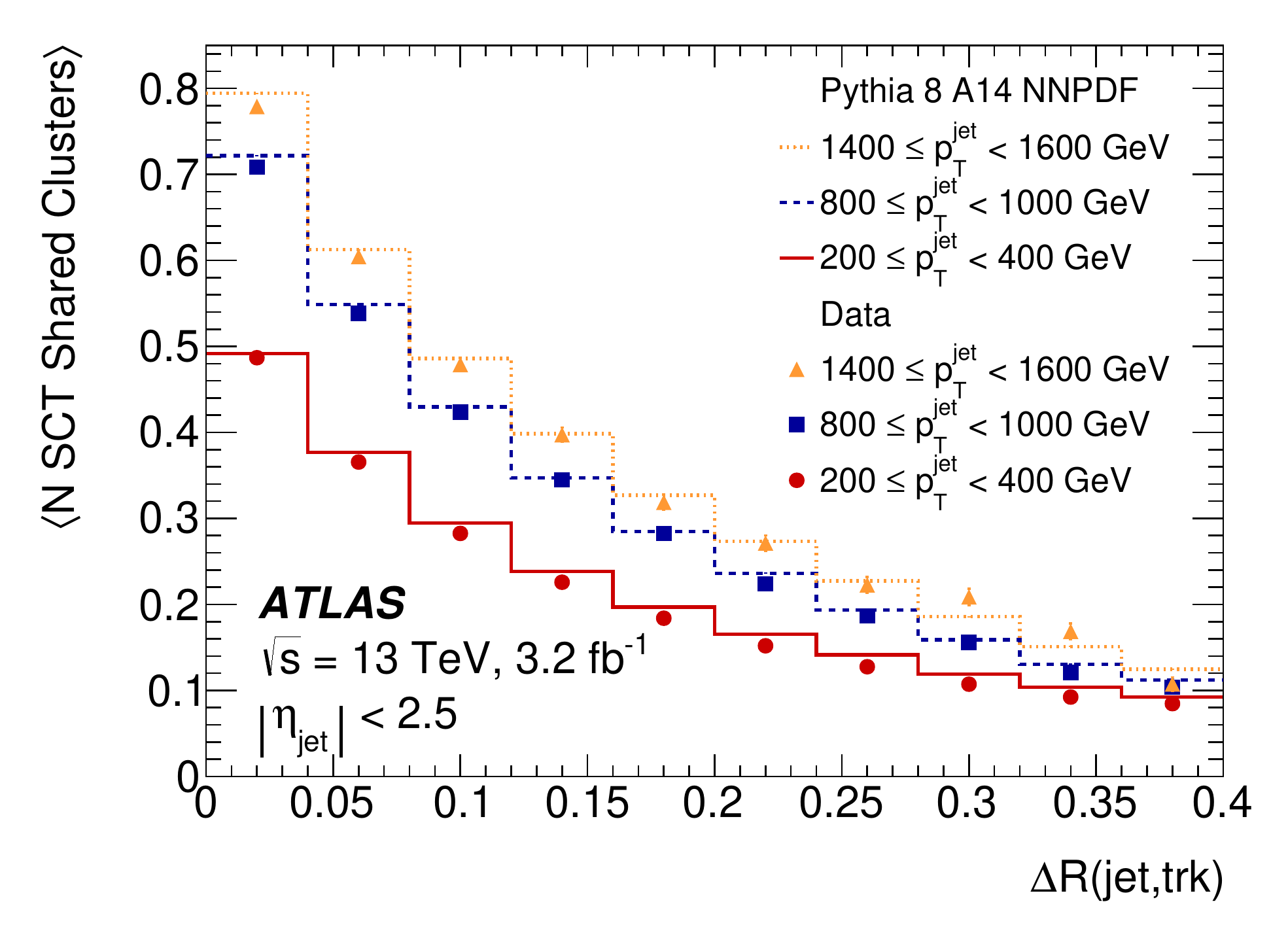}
\caption[Number of sct clusters on the track.]
{\label{fig:sctshared}
The average number of shared SCT clusters for primary tracks with a production vertex before 
the IBL is shown as a function of the angular distance of the track from
the jet axis. Data (markers) and dijet MC (lines) samples are compared in bins of jet \pT.
Due to the lack of charge information and the coarse sensor dimensions, the clusters cannot be readily identified as merged.
}
\end{figure}

\subsection{Performance for collimated tracks}
\label{subsec:singleparticles}

Quantities such as cluster assignment and track reconstruction efficiencies can be studied using truth information from simulation to elucidate the track reconstruction behaviour in the presence of highly collimated charged particles. This section utilizes the single-particle samples described in Section~\ref{sec:samples}. Figure~\ref{fig:avgMinSepSimple} shows how the minimum separation between charged particles at the IBL sensor surfaces evolves with the initial particle's \pT. For the same \pT, the density of the decay products may differ significantly: the lighter the initial particle, or the higher the multiplicity of its decay products, the smaller the distance. 
The degradation of the track reconstruction performance is mainly driven by the distance between charged particles and the charged-particle multiplicity in their vicinity. The results presented hereafter are therefore representative of the reconstruction performance in many physics processes, provided these parameters are known.
Throughout this section, unless otherwise noted, it is required that 
all charged particles are created before the IBL (production radius smaller than 29~mm) 
in all figures shown.

\begin{figure}[htbp]
\centering
\includegraphics[width=0.49\textwidth]{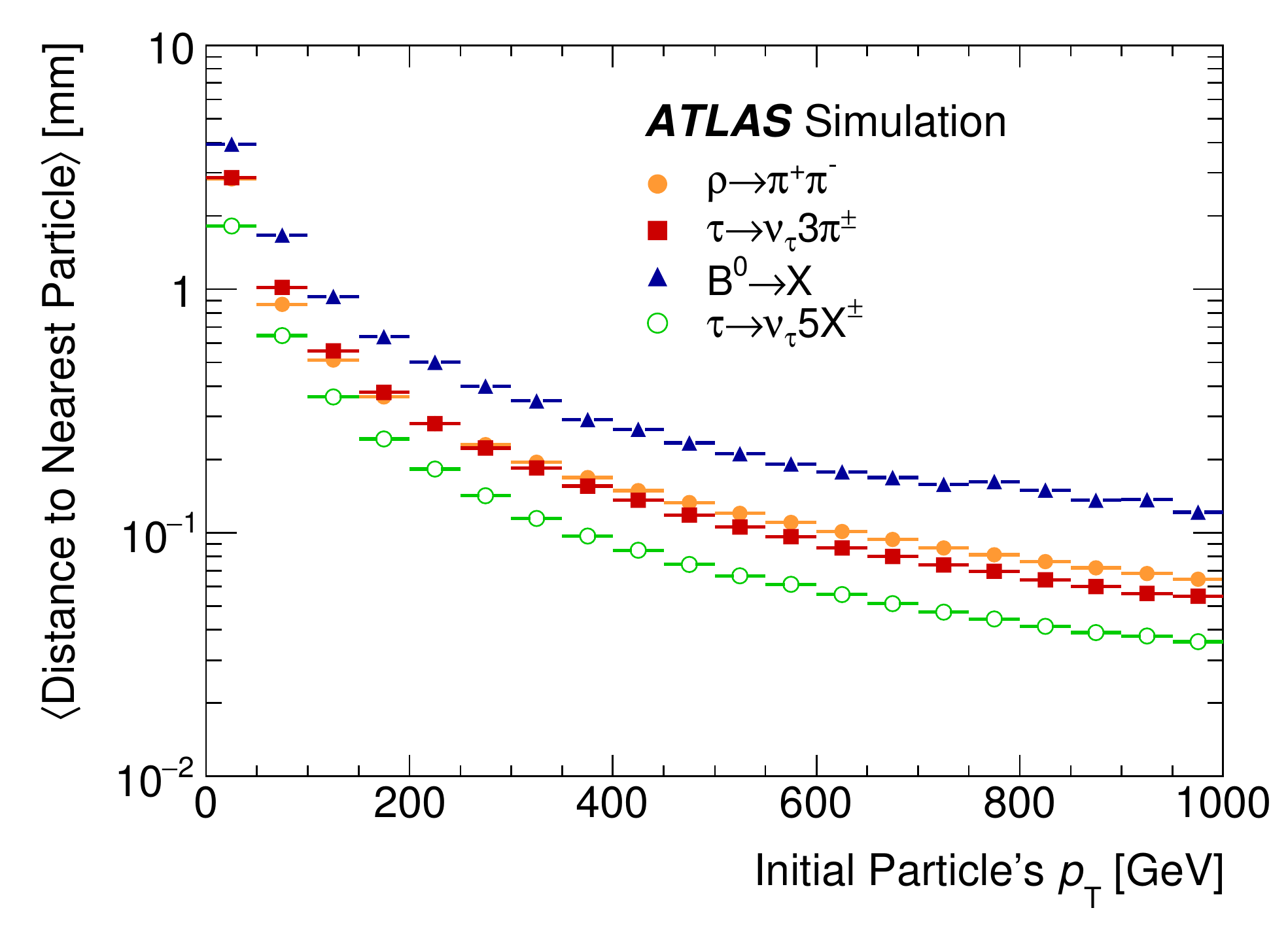}
\caption[Average minimum separation as a function of initial \pT\ for single-particle samples]
{\label{fig:avgMinSepSimple}
A comparison of the average minimum distance between charged decay products at the IBL sensor surfaces as a function of initial particle's \pT\ for single-particle samples.
}
\end{figure}

The average number of merged clusters 
is compared to the average number of clusters identified as merged 
in Figure~\ref{fig:SplitHit}
for the single $\rho$ and three-prong $\tau$ samples. 
The average charged-particle separation 
decreases with increasing initial-particle \pT\ leading to more
merged pixel clusters as shown in the points 
labelled \textit{Ideal}.
The average numbers of both the merged clusters and the clusters identified as merged fall to zero at the lowest initial-particle \pT, confirming a low rate of false-positives. Both grow at a similar rate with increasing initial-particle \pT. 
The residual inefficiency of the pixel NN is apparent in a lower number of 
clusters identified as merged compared to the ideal number of merged clusters at high initial-particle \pT.
The reconstruction performance correlates directly with the multiplicity and
distances at a given initial-particle \pT\ shown in Figure~\ref{fig:avgMinSepSimple}.

\begin{figure}[htbp]
\centering
\subfigure[$\rho\rightarrow\pi^{+}\pi^{-}$ sample] {\includegraphics[width=0.49\textwidth]{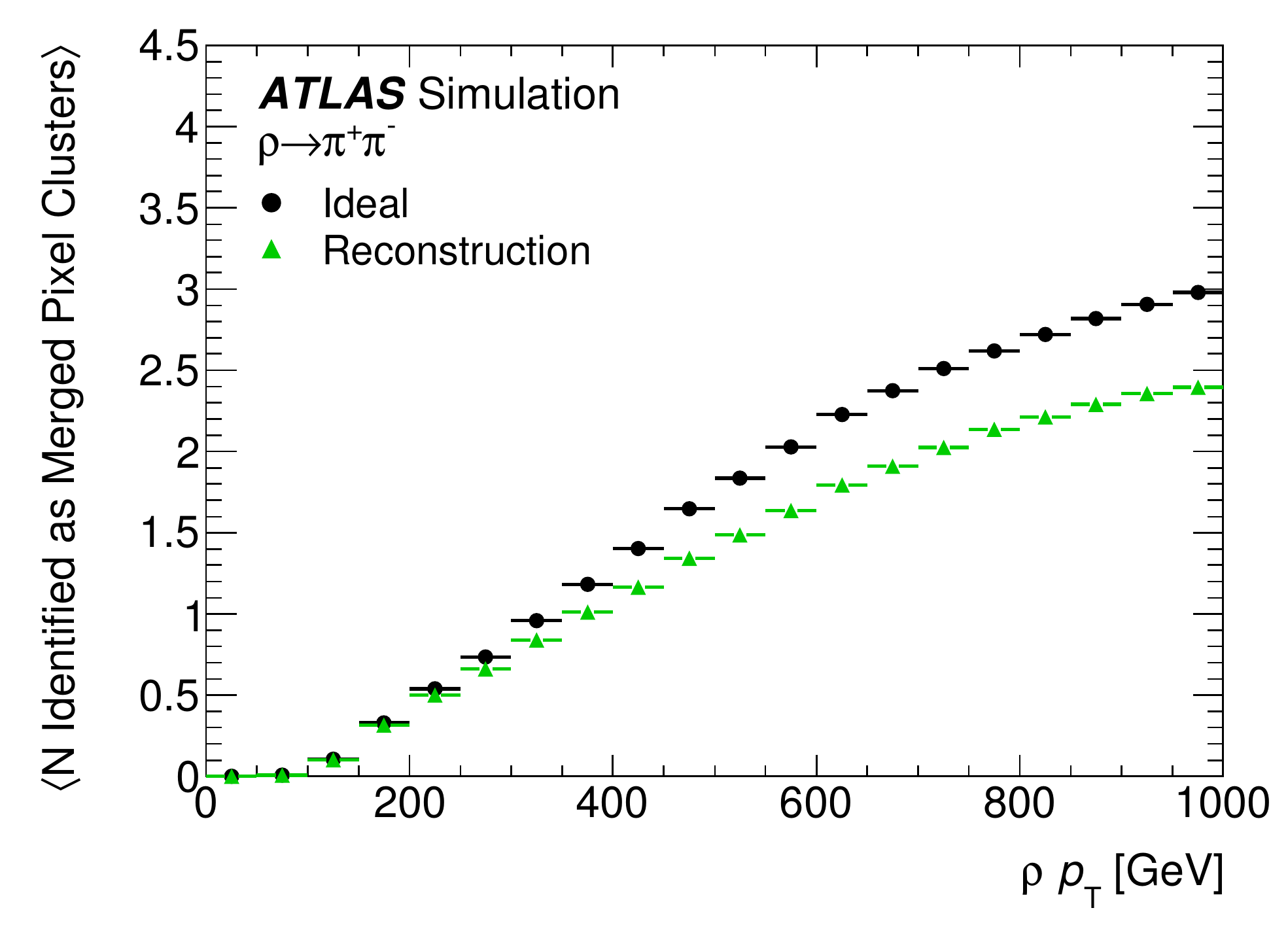}}
\subfigure[$\tau^{\pm} \rightarrow \pi^{+} \pi^{-} \pi^{\pm} \nu_{\tau}$ sample] {\includegraphics[width=0.49\textwidth]{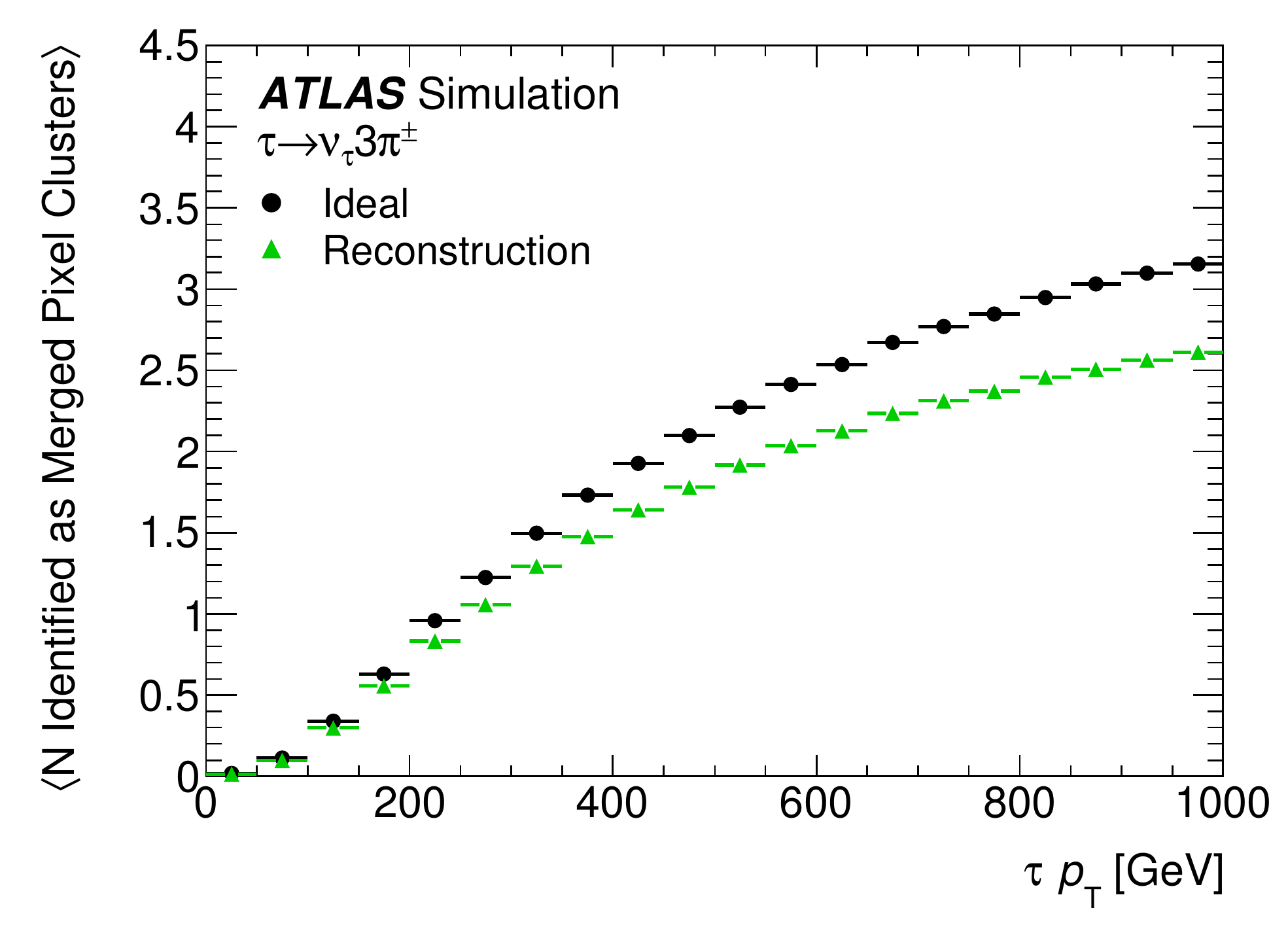}}
\caption[Merged/Split/Shareable clusters in $\rho$ and three-prong $\tau$ decays.]
{\label{fig:SplitHit}
A comparison of the average number of 
merged pixel clusters expected for truth particles from simulation
and pixel clusters identified as merged used in reconstructed tracks is shown as a function of the $\rho$ and three-prong $\tau$ ($\tau \rightarrow \pi^{+} \pi^{-} \pi^{\pm} \nu_{\tau}$) transverse momentum. Ideal represents the true number of merged clusters, which would be obtained as the number of identified merged clusters in the case of perfect performance. It is required that the stable charged particles are created before the IBL.
}
\end{figure}

Merged clusters failing identification can result in shared clusters, which (as explained in Section~\ref{sec:ambisolving}) need to be limited. To study possible inefficiencies of the reconstruction algorithm, the cluster assignment efficiency is shown in Figure~\ref{fig:hitEffeta} as a function of the minimum truth particle separation at the sensor's surface for the first two layers of the pixel detector. It is defined as the fraction of 
clusters created by a particle that are then used on the reconstructed track of said particle.
With the closest truth particle separated by
400~$\mu$m at the IBL, the cluster assignment efficiency
at this layer is in excess of 99\% for the 
$\rho$ and three-prong $\tau$ samples, and 98\% for the $B^{0}$ samples.
When going to smaller separations, individual clusters start to merge and eventually only a single merged cluster remains. Since in the simpler topology $\rho\rightarrow\pi^{+}\pi^{-}$ the cluster has to be assigned to a maximum of two tracks, the cluster assignment efficiency is 99\% down to the smallest distances shown. 
In case of the $B^{0}$ and three-prong $\tau$ decays, several daughter particles are likely to contribute
to a merged cluster. The NN described in Section~\ref{subsec:pixelclustering} lacks the ability to distinguish between merged clusters from more than three particles and those from exactly three particles~\cite{PERF-2012-05}. Also, the track reconstruction algorithm limits the number of tracks using the same cluster without penalties to three. As a result, at much smaller particle separations, the cluster assignment efficiency is limited in the $B^{0}$ and three-prong $\tau$ samples. The case of more than three charged particles contributing to a pixel cluster in the $B^{0}$ decay results in an additional assignment inefficiency on the B-layer.  

\begin{figure}[htbp]
\subfigure[$\rho\rightarrow\pi^{+}\pi^{-}$ sample] {
  \includegraphics[width=0.49\textwidth]{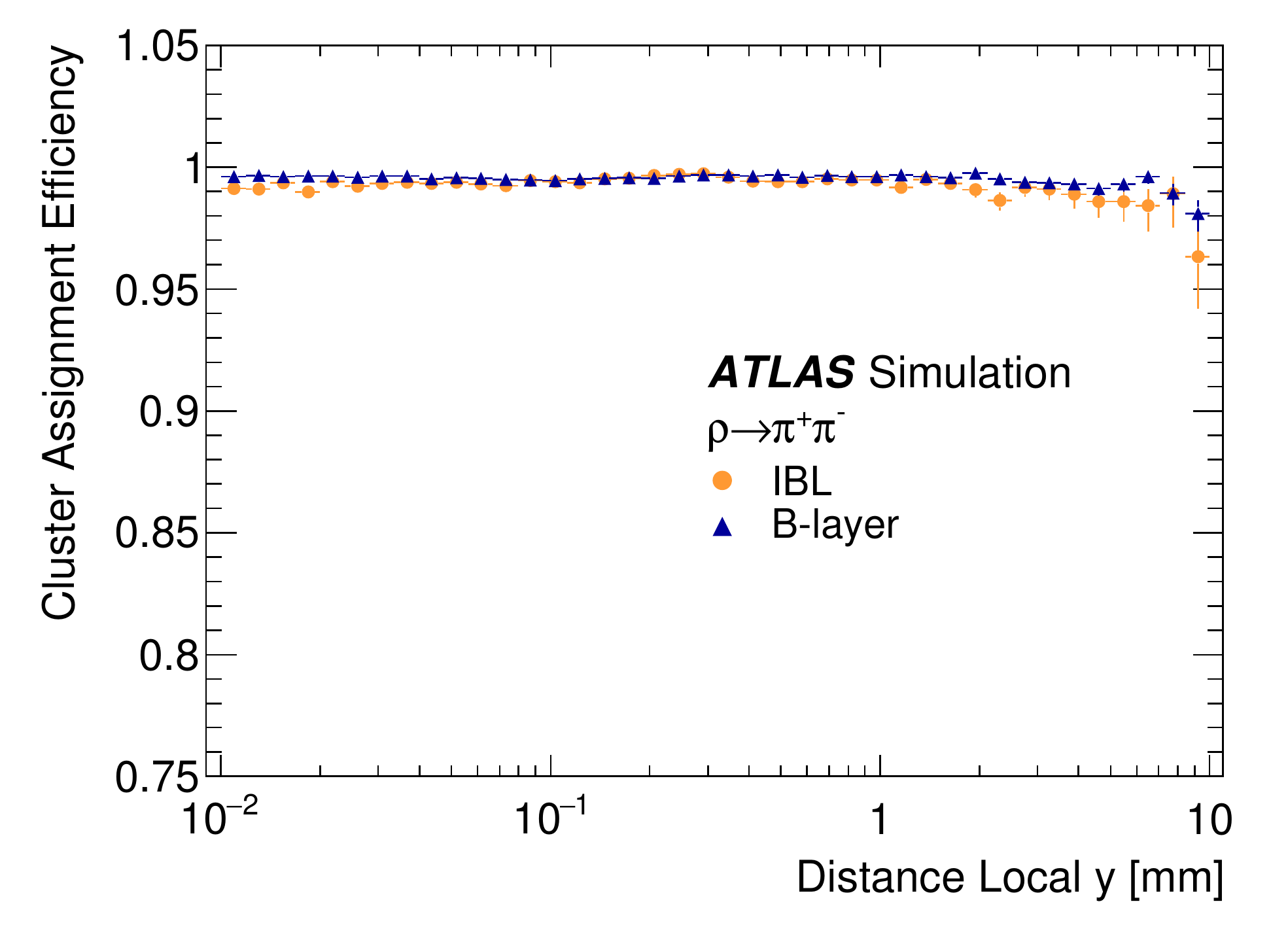}
}
\subfigure[$\rho\rightarrow\pi^{+}\pi^{-}$ sample] {
  \includegraphics[width=0.49\textwidth]{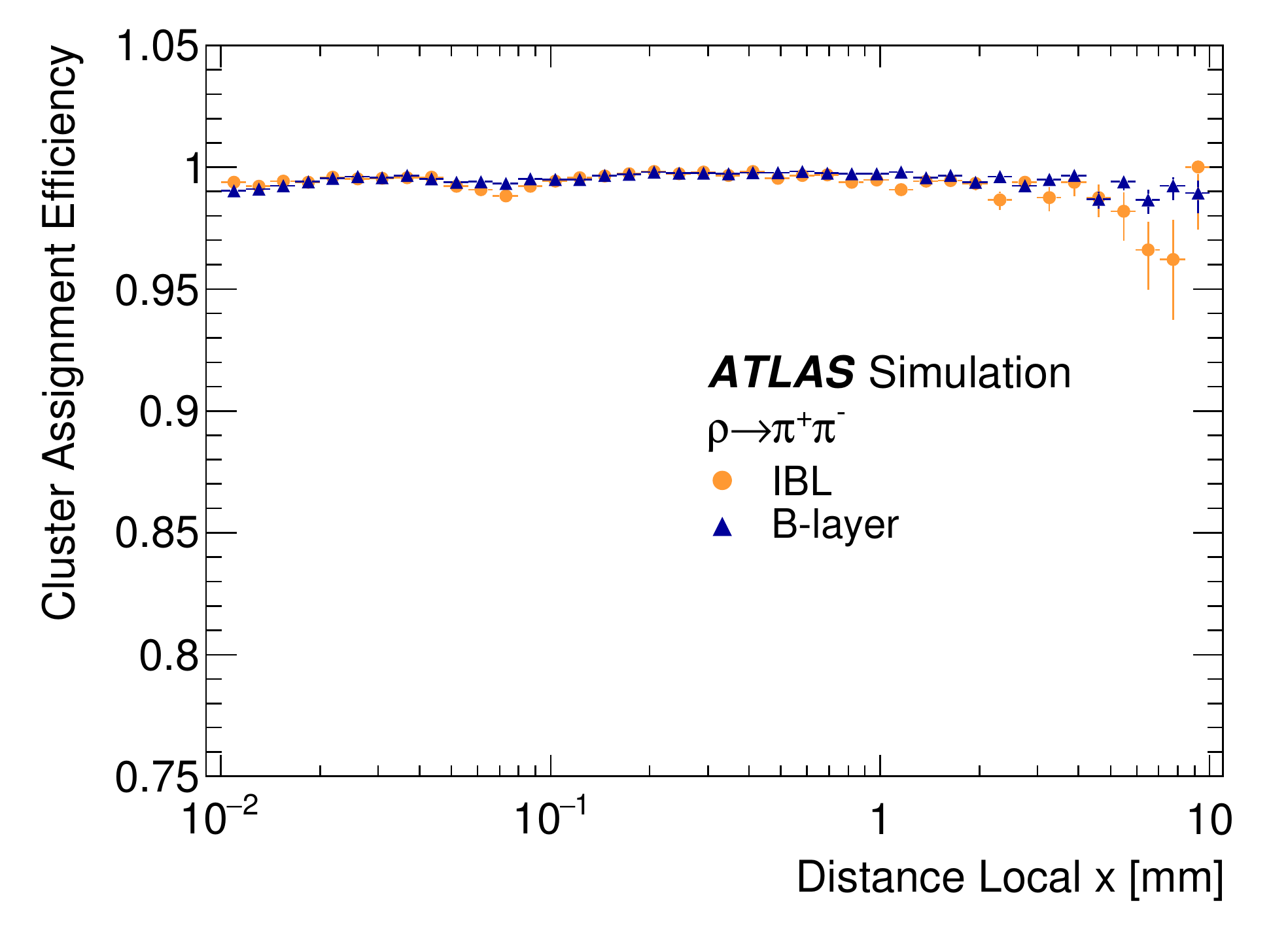}
} \\
\subfigure[$\tau\rightarrow \pi^{+} \pi^{-} \pi^{\pm} \nu_{\tau}$ sample] {
  \includegraphics[width=0.49\textwidth]{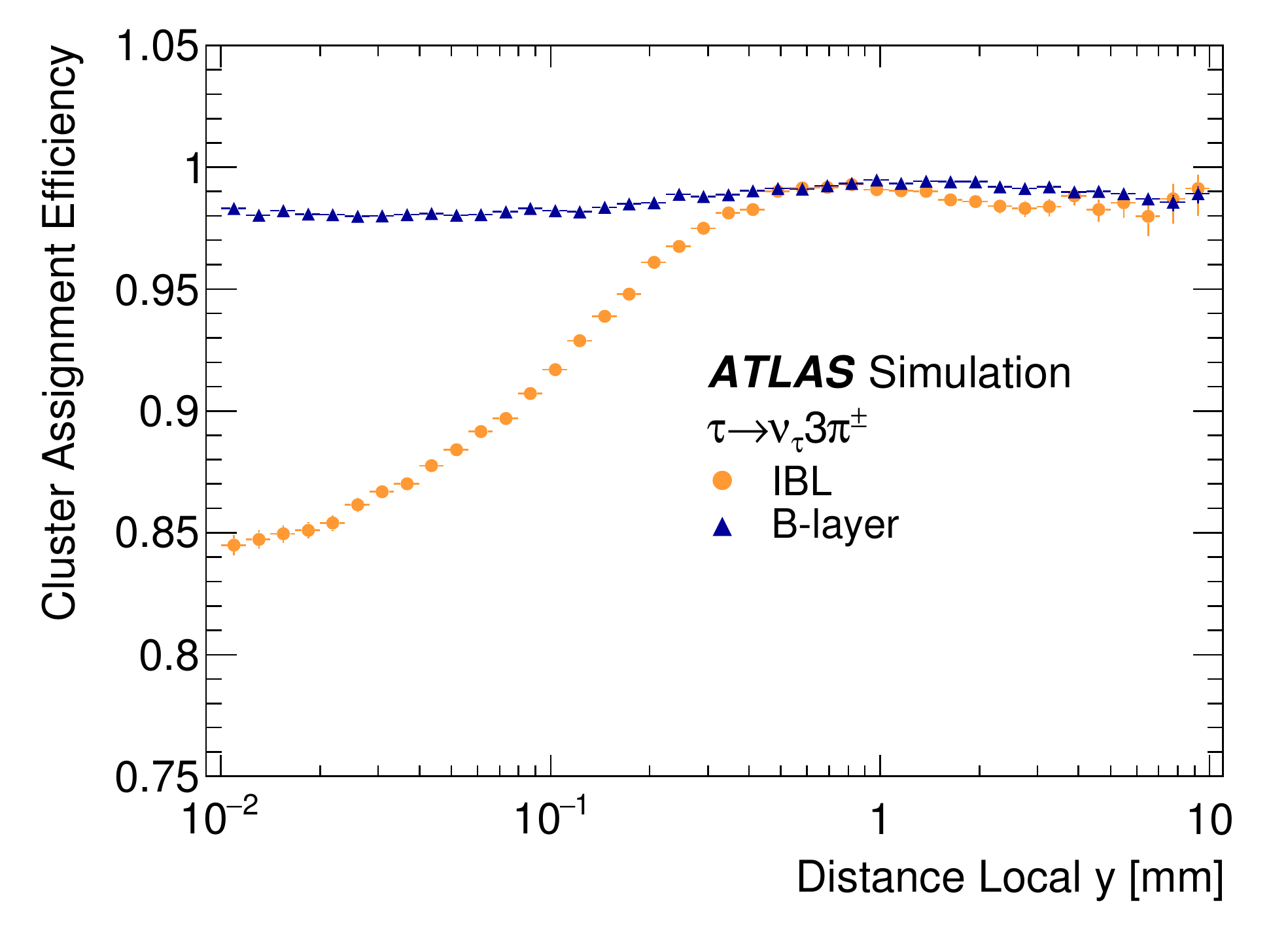}
}
\subfigure[$\tau\rightarrow \pi^{+} \pi^{-} \pi^{\pm} \nu_{\tau}$ sample] {
  \includegraphics[width=0.49\textwidth]{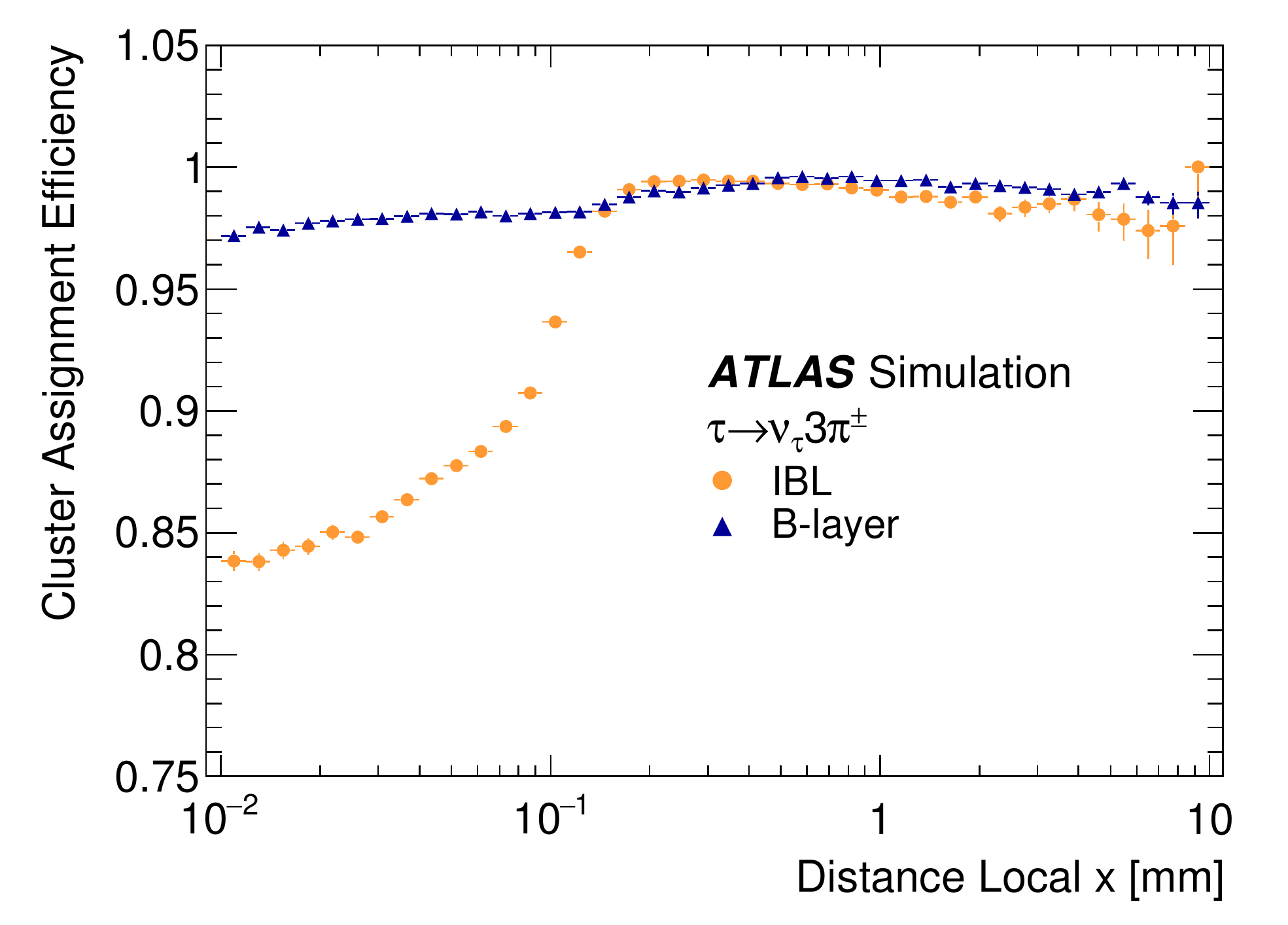}
} \\
\subfigure[$B^{0} \rightarrow X$ sample] {
  \includegraphics[width=0.49\textwidth]{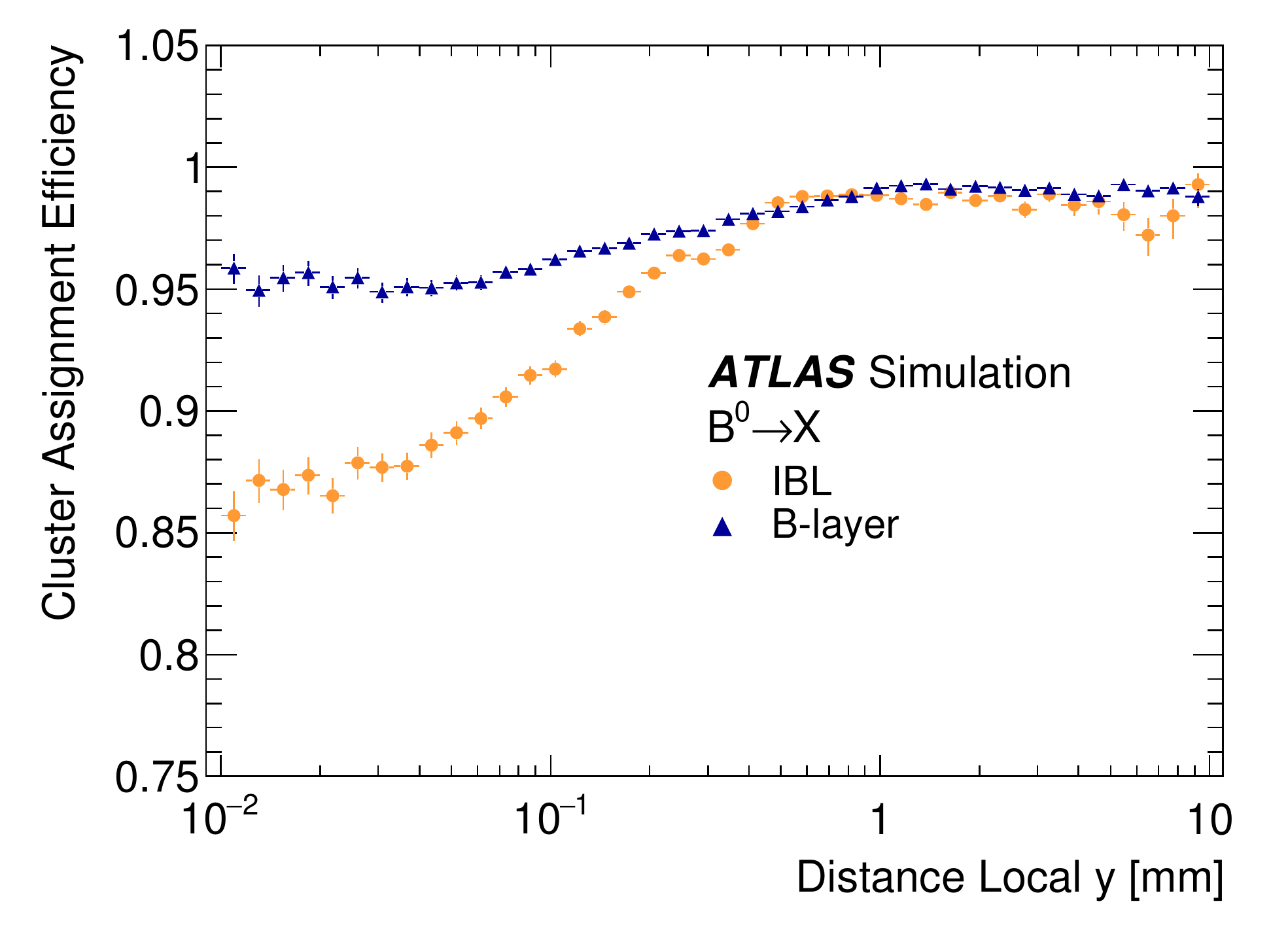}
}
\subfigure[$B^{0} \rightarrow X$ sample] {
  \includegraphics[width=0.49\textwidth]{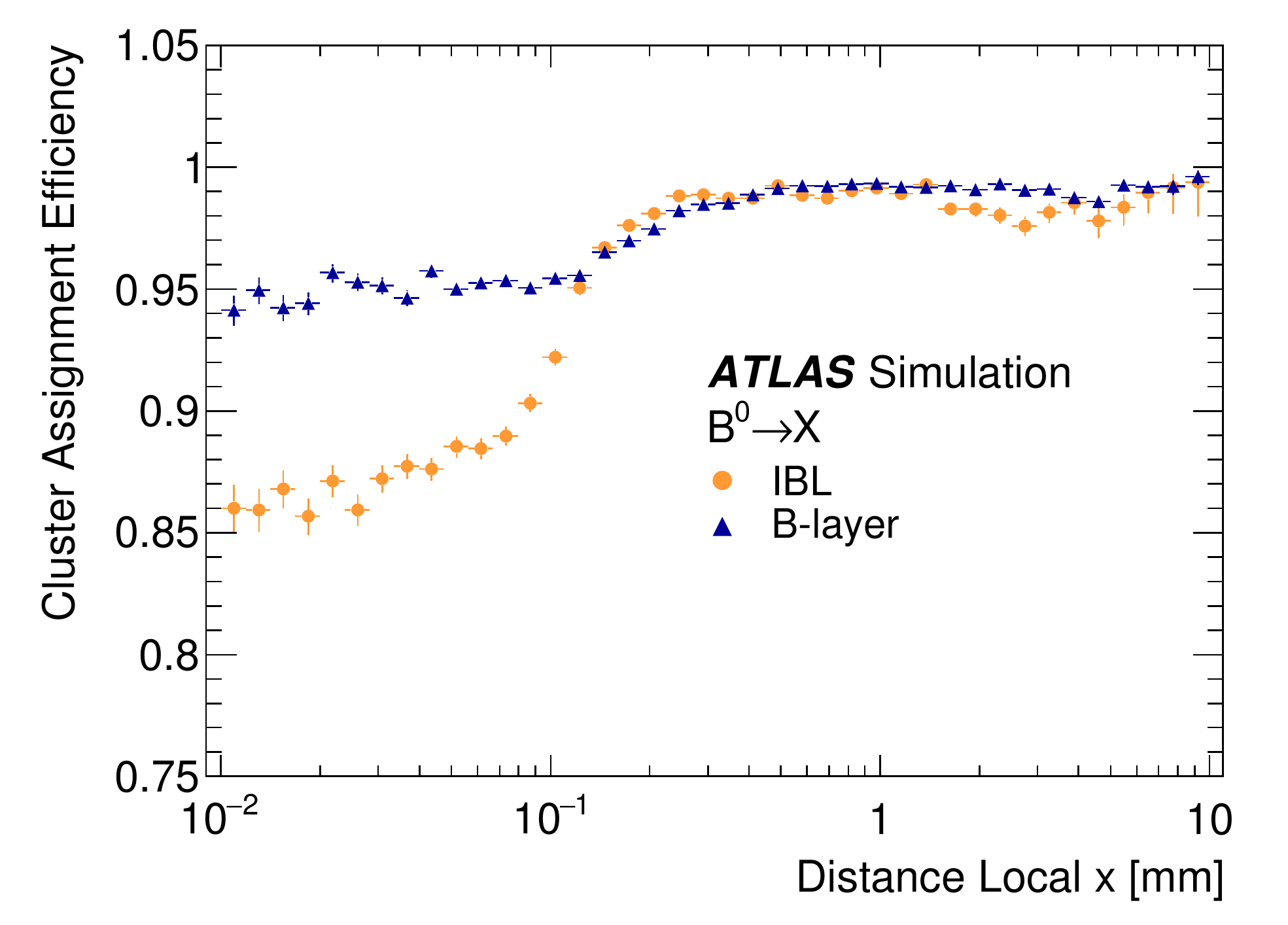}
}
\caption[Cluster assignment efficiency.]{\label{fig:hitEffeta} 
For the $\rho$ (top), three-prong $\tau$ (middle), and $B^{0}$ (bottom) samples,
the efficiency with which reconstructed clusters are properly assigned to a track is
shown for the two innermost pixel layers (IBL and B-layer) as a function 
of the minimum truth-particle separation in local $y$ (left) and $x$ (right), 
corresponding to the pixel dimensions longitudinal and transverse to the beam axis. It is required that the stable charged particles are created before 
the IBL.
}
\end{figure}

Regardless of how well the ambiguity solver identifies merged pixel clusters and assigns them to tracks, a substantial inefficiency remains at high initial-particle momenta due to the necessary limitations on shared SCT clusters. 
Figure~\ref{fig:tauEffShared} shows the \textit{reconstructable efficiency} of the $\rho$ and three-prong $\tau$ decays utilizing MC truth information. This is defined as the efficiency to be able to reconstruct all of the charged decay products from a given resonance having satisfied the cluster multiplicity requirements defined in Section~\ref{sec:ambisolving}.
All merged pixel clusters are assumed to have been identified, so for a fixed maximum number of allowed shared SCT clusters, this represents the maximum achievable reconstruction efficiency. The loss in efficiency is exacerbated by increasing charged-particle multiplicities as in the three-prong $\tau$ sample.
This limit is fixed at two shared clusters. The 
efficiency improvement obtained from loosening this limit
is not sufficient to justify the associated increase in the proportion of fake tracks. In simulated events with several jets, the inclusive number of fake tracks increases by 25\% when loosening the limit to three shared clusters.

\begin{figure}[htbp]
\centering
\subfigure[$\rho\rightarrow\pi^{+}\pi^{-}$ sample] {\includegraphics[width=0.49\textwidth]{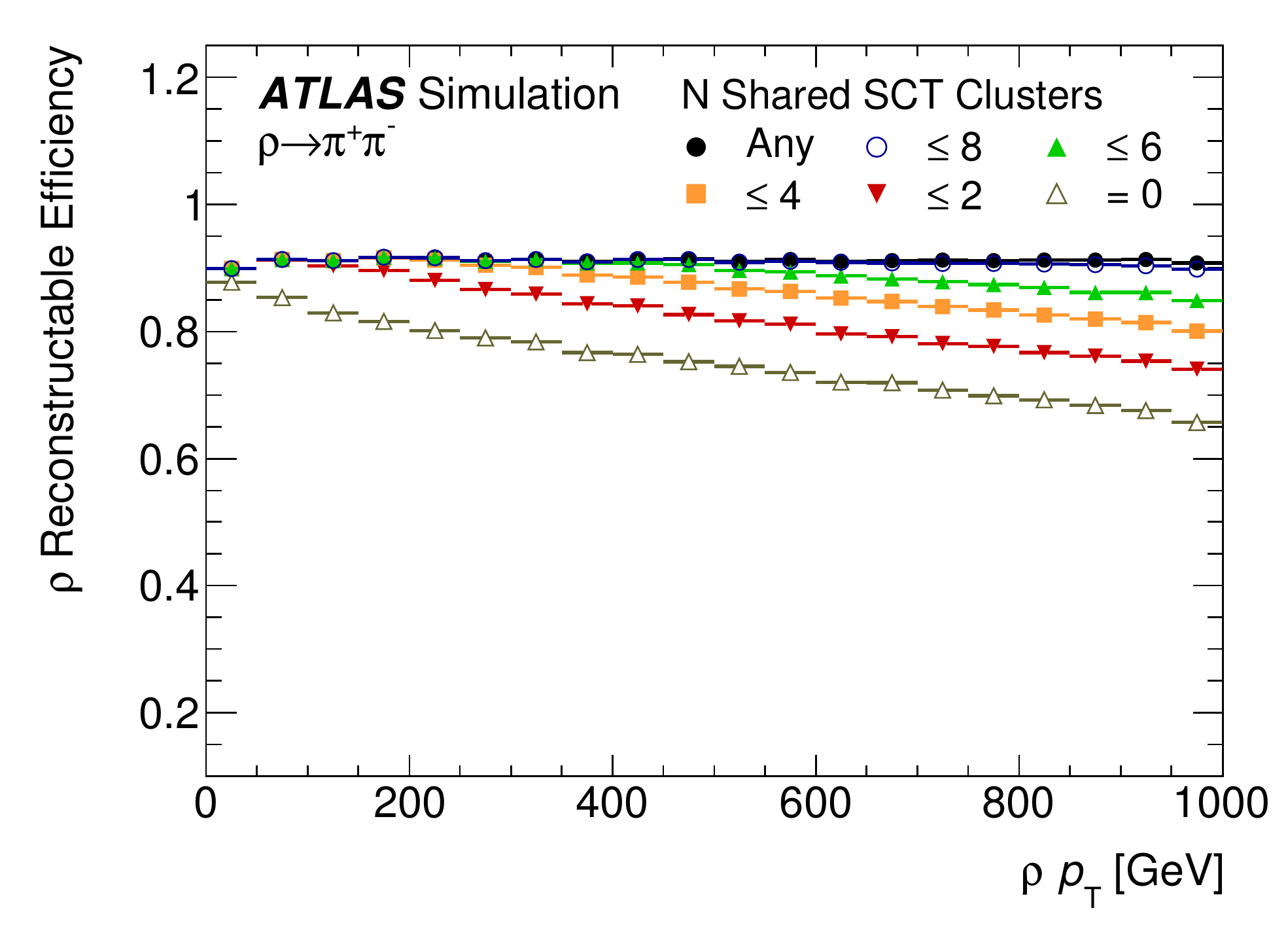}}
\subfigure[$\tau^{\pm} \rightarrow \pi^{+} \pi^{-} \pi^{\pm} \nu_{\tau}$ sample] {\includegraphics[width=0.49\textwidth]{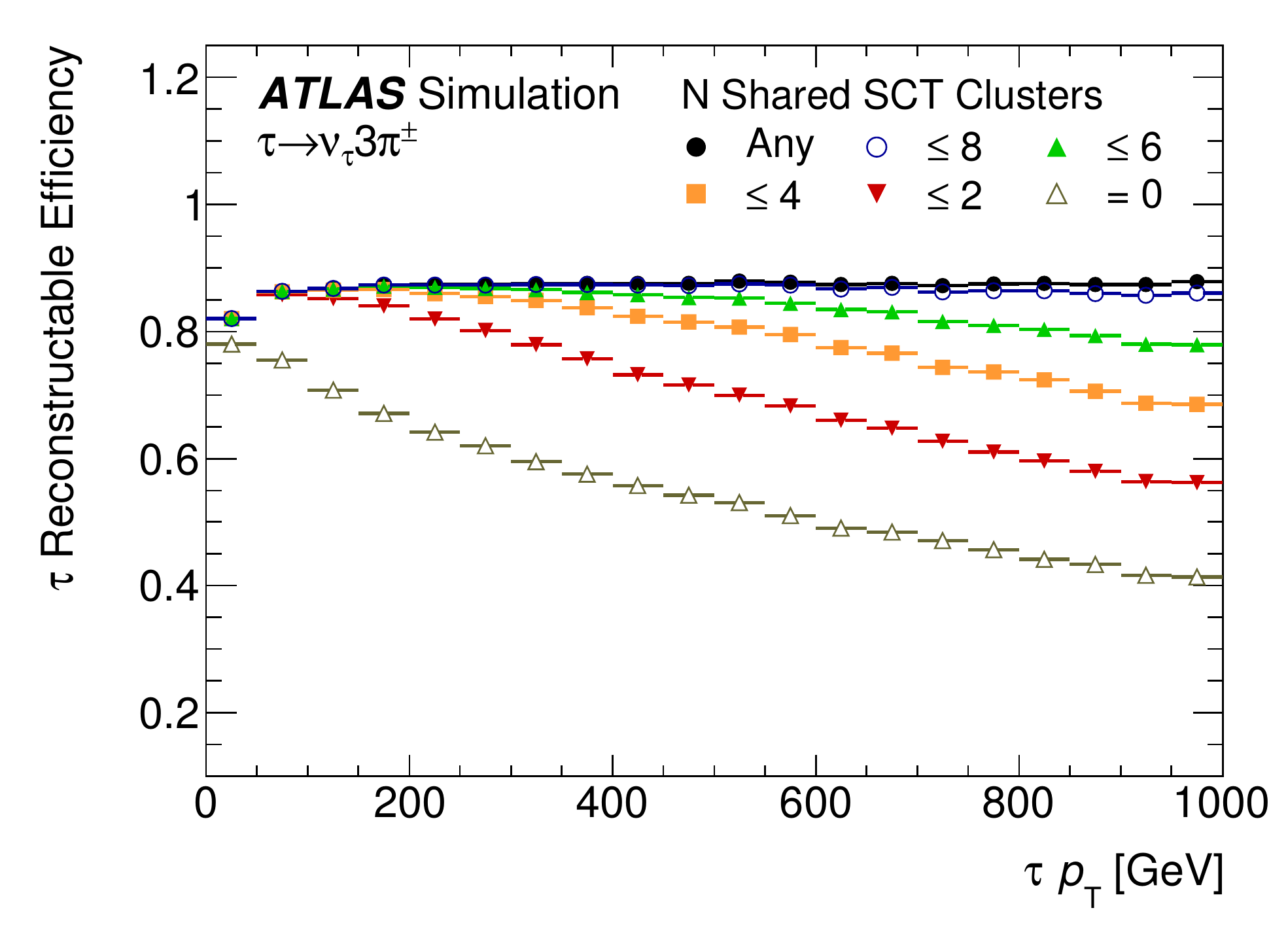}}
\caption[Shared cluster cut impact on maximum efficiency for 
$rho$s and three prong $\tau$s.]{\label{fig:tauEffShared} 
The reconstructable efficiency, defined as the efficiency to reconstruct all of the charged decay products of the parent particle, is shown
for the $\rho$ and three-prong $\tau$ samples
with various limits on the number of shared clusters allowed on a track candidate
assuming all the merged pixel clusters have been identified as merged. It is required that the stable charged particles are created before the IBL.
}
\end{figure}

Finally, the per-track reconstruction efficiency is shown in Figure~\ref{fig:tauRecEffNoShared}
as a function of particle \pt and production radius. The production radius is defined as the radial distance of the decay of the parent particle from the beam axis.
The efficiency degrades with increased multiplicity.
The visible inefficiency in all samples at low initial-particle \pt is due to inelastic interactions, 
such as hadronic interactions. At higher transverse momentum of the initial particle, a decrease in efficiency is driven by the increasingly collimated nature of the decay products. 
A decrease in efficiency is also seen with a increasing production radius 
as the charged particles arrive at each active layer with less average separation.
The requirement on the total number of clusters for track reconstruction
leads to discrete drops in efficiency at each active layer.

\begin{figure}[htbp]
\centering 
\subfigure[Efficiency versus initial particle's \pt] {\includegraphics[width=0.49\textwidth]{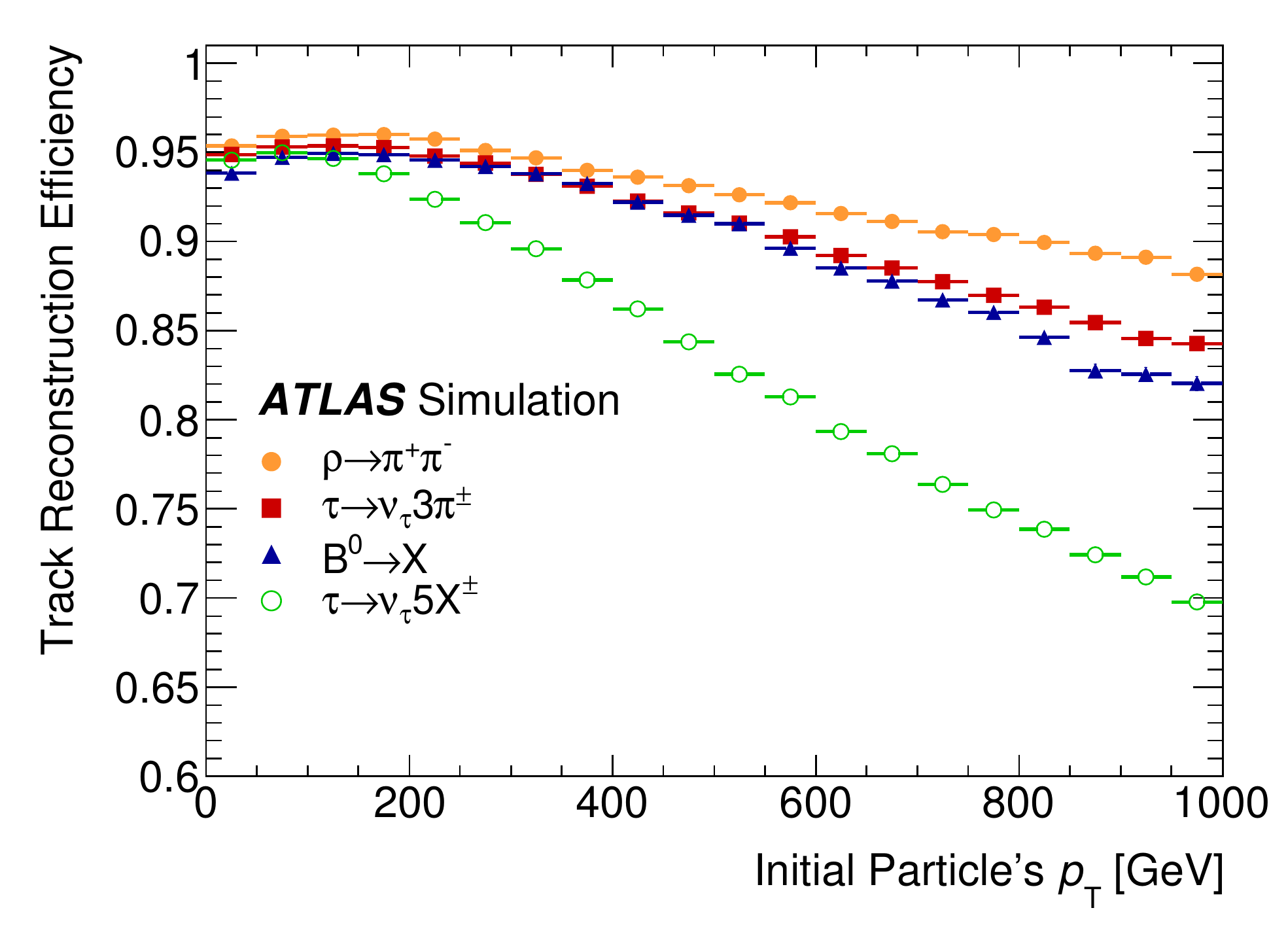}}
\subfigure[Efficiency versus production radius] {\includegraphics[width=0.49\textwidth]{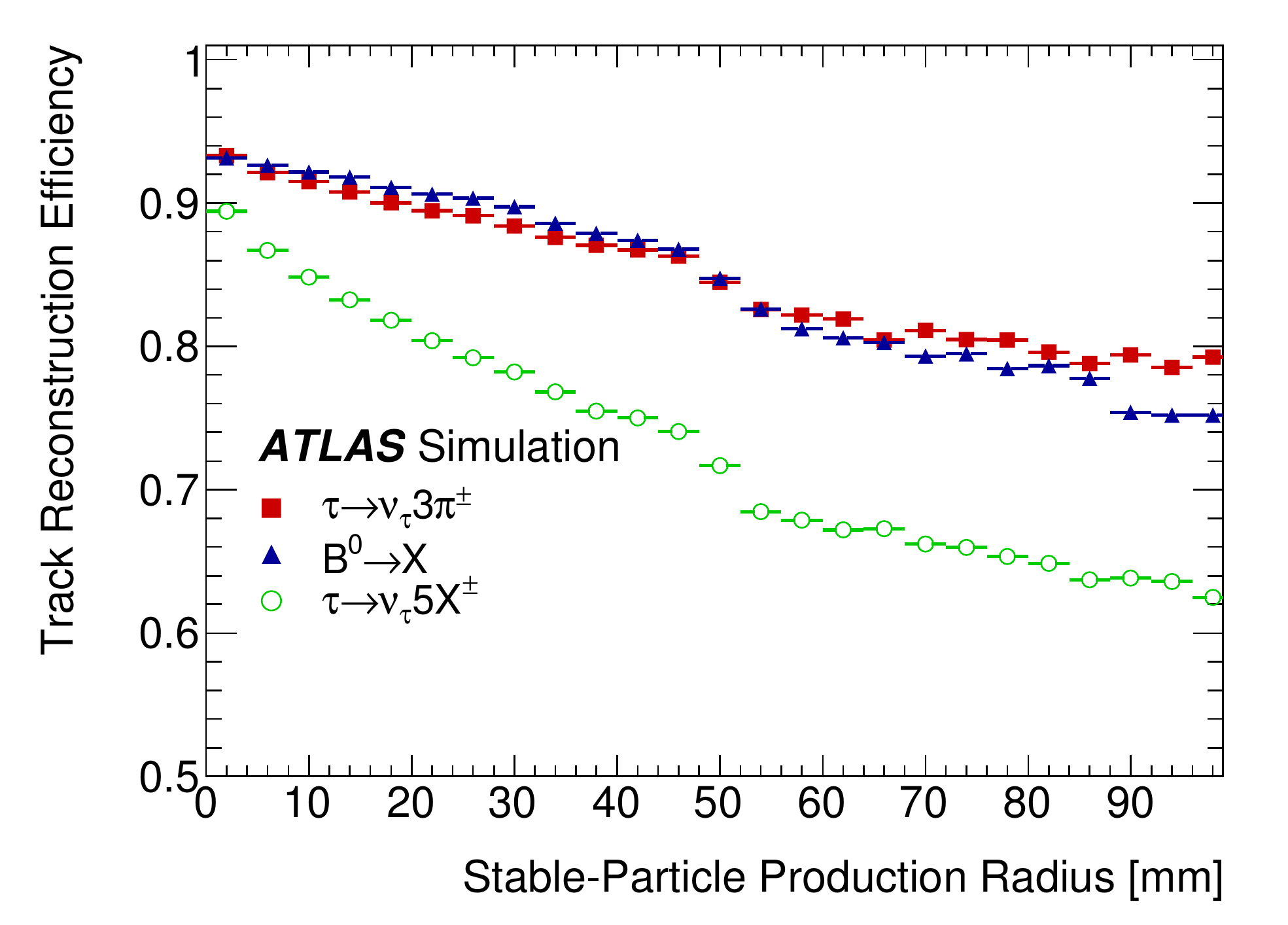}}
\caption[Reconstruction efficiency]{\label{fig:tauRecEffNoShared} 
Single-track reconstruction efficiency is shown as
(a) a function of the initial particle's \pT\ when it is required that the 
parent particle decays before the IBL 
for the decay products of a $\rho$, three- and five-prong $\tau$ and a $B^{0}$ 
and, (b) versus the production radius
for the decay products of a three- and five-prong $\tau$ as well as a $B^{0}$, where no requirement is imposed on the production radius of stable charged particles.
}
\end{figure}

\subsection{Performance for tracks in jets}
\label{subsec:jets}

In the previous sections, the performance in simple topologies is discussed. These samples are crucial for understanding the effects of charged-particle separations and multiplicities on the performance, but they are insufficient to quantify the expected performance in the dense jet environments evident in Figure~\ref{fig:jetpt}. As demonstrated in Section~\ref{sec:basicdatamc}, samples of dijet MC events do provide a reasonable description of jets in data. The following contains studies of the track reconstruction efficiency in these samples.

Figure~\ref{fig:EffLowHigh} shows the charged-primary-particle 
reconstruction efficiency dependence on the angular distance of 
a particle to the jet axis for different jet $\eta$ and \pT ranges. All charged particles studied are required to be created before the IBL. 
The efficiency drops rapidly towards the centre of the jet, where
the charged-particle density is maximal. 
A slight decrease in efficiency towards the edge of the jet is
consistent with an isolated-track efficiency that rises with charged-particle \pT~\cite{STDM-2015-17}
and a decrease in the average charged-particle \pT with distance from the jet core.
The dependence of the efficiency on the jet \pT\ and on the production radius of the charged particle, where charged particles are not required to be created before the IBL, is shown in Figure~\ref{fig:EffAllEta}. The decrease in efficiency with production radius is from two effects. Firstly, particles created beyond the first active layers of the ID create fewer clusters. Secondly, with the shorter flight length to the next active layer, the average separation between particles is smaller compared to prompt decays, producing more merged clusters.
The overall trend for all efficiencies shown is the same at all $\eta$. However, the loss in absolute efficiency is exacerbated at high $|\eta|$, while the degradation at small separations between a track and the jet axis is alleviated.

\begin{figure}[htbp]
\centering
\subfigure[]{\includegraphics[width=0.49\textwidth]{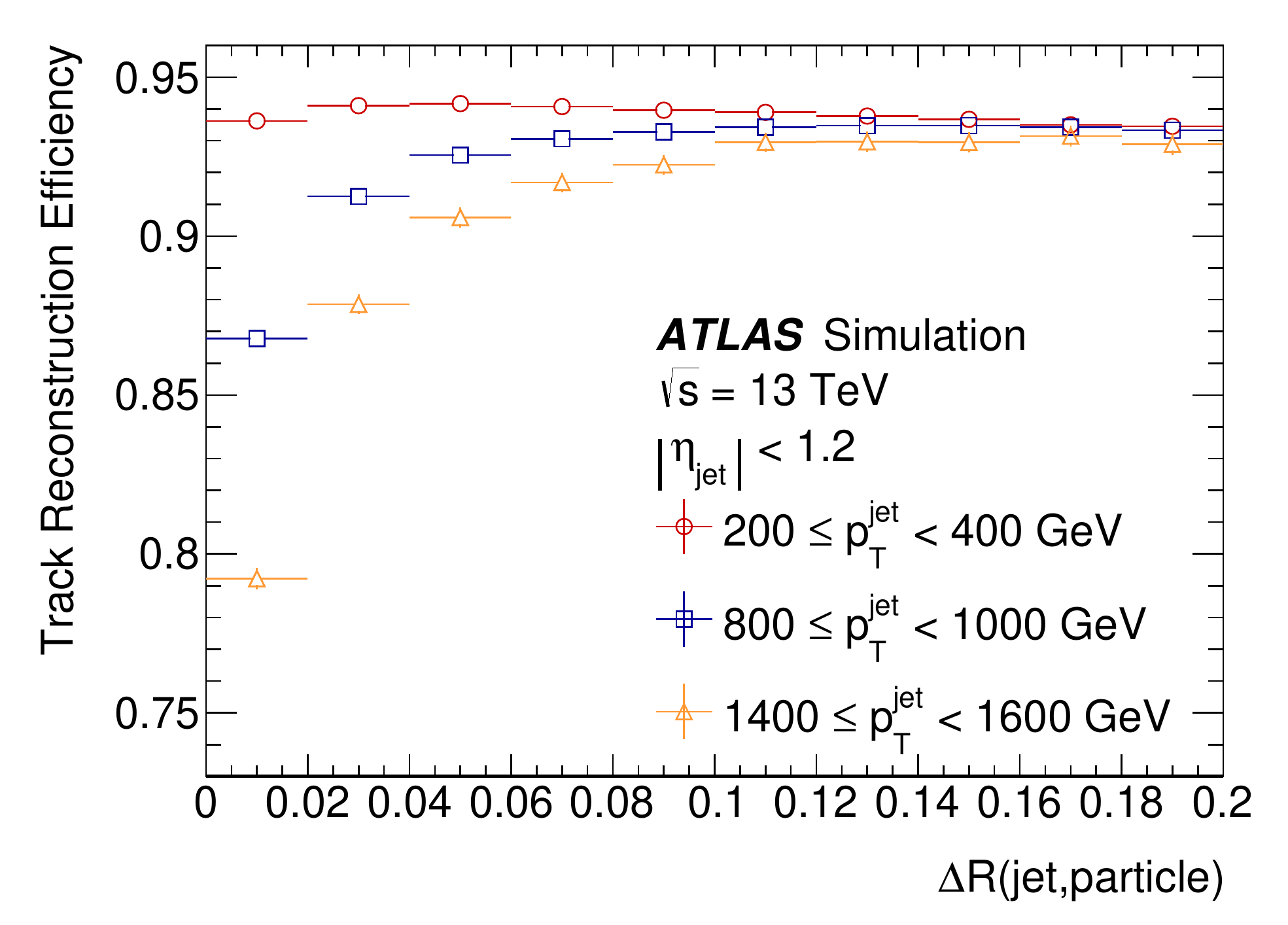}}
\subfigure[]{\includegraphics[width=0.49\textwidth]{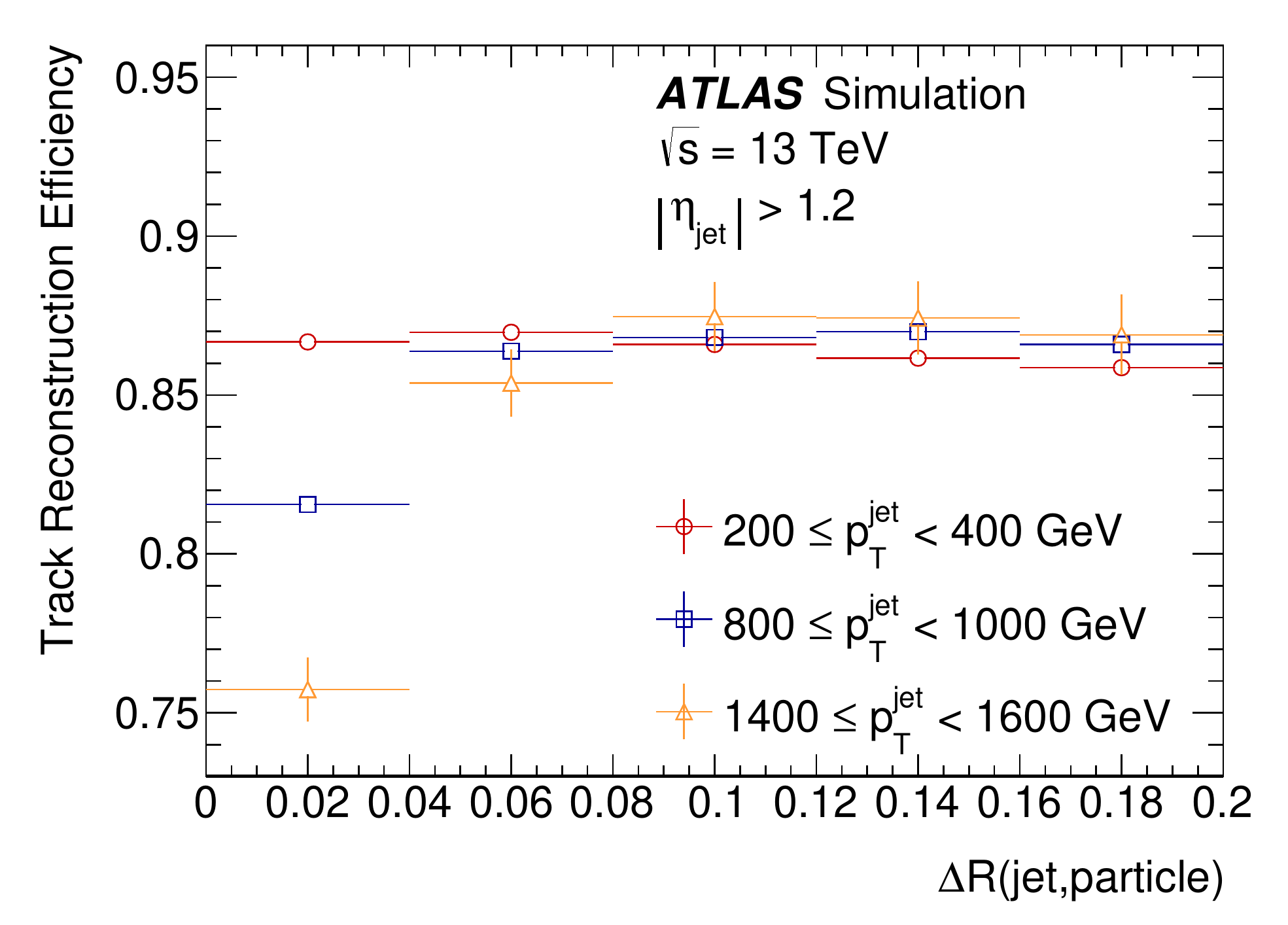}}
\caption[CMS comparison plots.]{\label{fig:EffLowHigh} 
  The efficiency to reconstruct charged primary particles in jets with (a) $|\eta|<1.2$ and (b) $|\eta|>1.2$ is shown 
as a function of the angular distance of the particle from the jet axis for various jet \pT for simulated dijet MC events.
}
\end{figure}

\begin{figure}[htbp]
\centering
\subfigure[]{\includegraphics[width=0.49\textwidth]{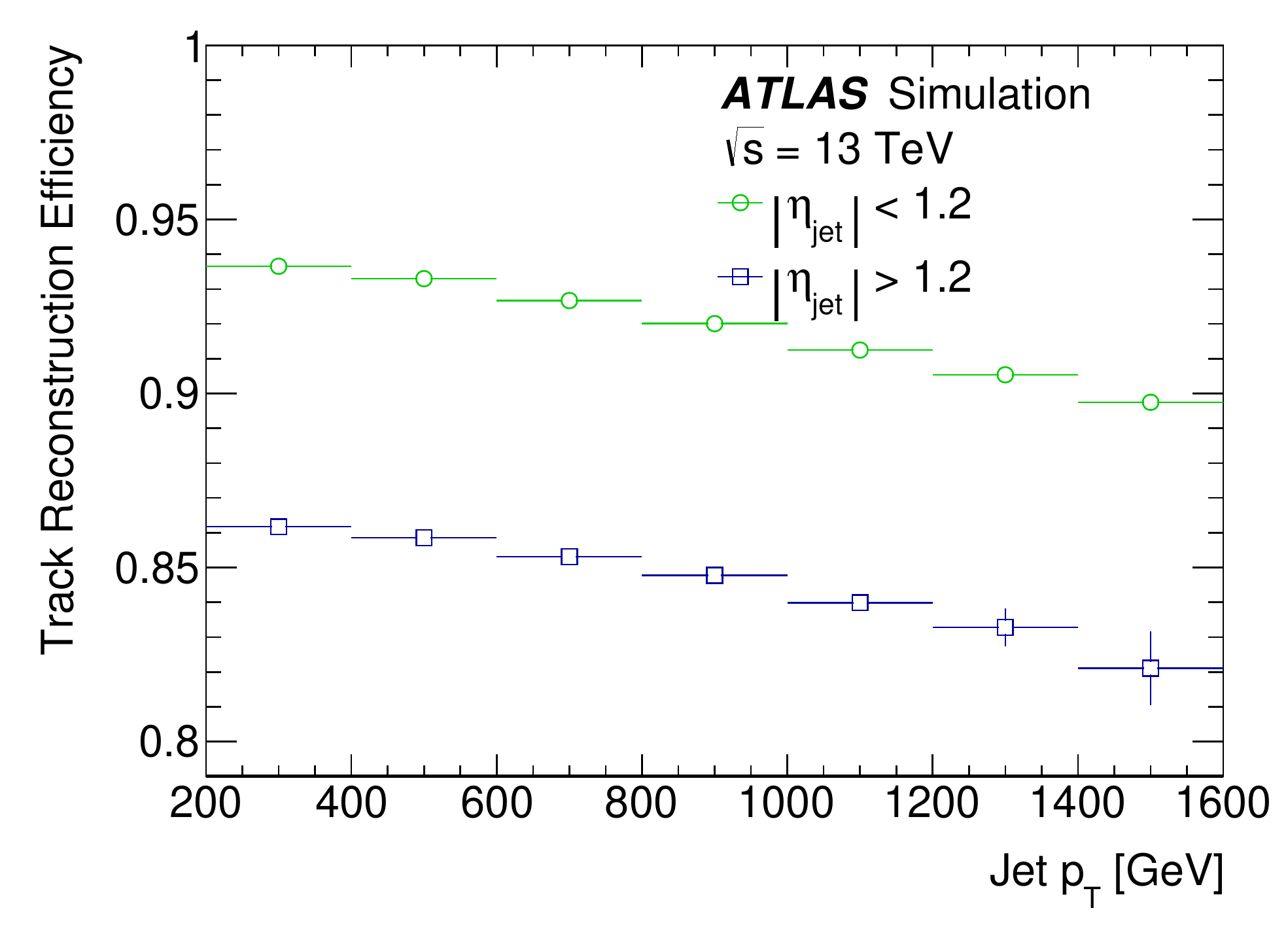}}
\subfigure[]{\includegraphics[width=0.49\textwidth]{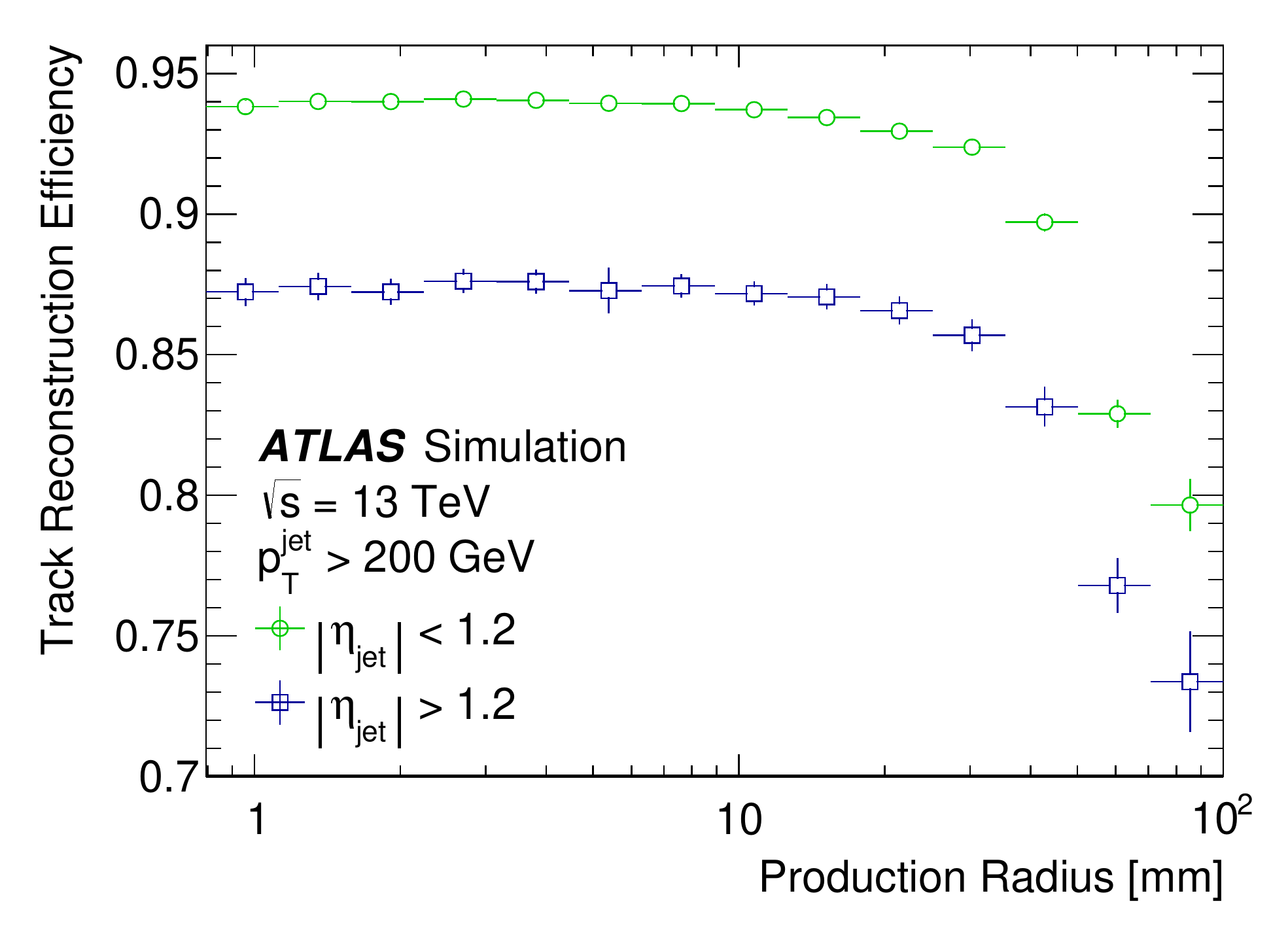}}
\caption[CMS comparison plots.]{\label{fig:EffAllEta} 
The track reconstruction efficiency is compared for charged primary particles in jets with $|\eta|<1.2$ ($|\eta|>1.2$) for the entire jet-\pT range as a function of 
(a) the jet \pT and
(b) the production radius of the charged particle for simulated dijet MC events, where charged particles are not required to be created before the IBL.
}
\end{figure}

\FloatBarrier
\section{Measurement of track reconstruction efficiency in jets from data}
\label{sec:dEdx}
Previous sections discuss the performance of the track reconstruction in dense environments based mainly on MC simulation. This section introduces a novel method to probe this performance in data. A measurement of the fraction of tracks lost in reconstruction due to the high density and collimation of charged particles in high-\pT jets is presented for the subset of tracks with a B-layer cluster created by two charged particles. 

The \dEdx\ of a charged particle traversing the pixel sensor is measured from the charge collected in the clusters associated with the reconstructed track. With single particles and thin layers, one expects the \dEdx\ measurements to approximately follow a Landau distribution~\cite{bib:PDG}. A typical particle reconstructed from an LHC collision is expected to be a minimum-ionizing particle (MIP). Thus, two particles contributing to the same cluster are expected to deposit twice the energy of a single MIP. In the context of this paper, \dEdx\ is normalized to the material density, and it therefore has units of \dEdxUnit.

As demonstrated in the previous sections, near the jet core the charged-particle density is high and particles can be highly collimated. The tracks of these particles are thus more likely to create merged clusters, as shown in Figure~\ref{fig:NPixOnTrk}. 
By fitting the cluster~\dEdx~for reconstructed tracks near the core of the jet, single-particle clusters can be statistically separated from merged clusters. The fraction of lost tracks can therefore be inferred from the number of times only one reconstructed track is associated with a cluster \dEdx\ compatible with two MIPs. At truth-level, this fraction is defined as follows: the denominator is the number of truth particles passing the analysis selections (listed in Section~\ref{subsec:selections}, and including a $p_{\textrm{T}}>10$~\GeV\ requirement), which have a B-layer cluster created by exactly two charged particles; the numerator is the subset of these particles which failed to be reconstructed.

For the IBL, ToT is encoded in four bits. Eight bits are available in each of the remaining three pixel layers, which therefore provide an enhanced ToT resolution compared to the IBL, resulting in a superior energy resolution. For this reason, the cluster \dEdx\ values corresponding to the B-layer are used in this study.

\subsection{Track selection}
\label{subsec:selections}
To enhance the contribution of high-quality collimated tracks and suppress fake tracks to a negligible number, additional track selections beyond those outlined in Section~\ref{sec:ambisolving} are required for all tracks used in this analysis:
\begin{itemize}
\item Exactly one pixel cluster per layer,
\item $p_{\textrm{T}} >$ 10~\GeV,
\item $|\eta| <$ 1.2,
\item $|d_0^{\textrm{BL}}| <$ 1.5 mm,
\item $|z_0^{\textrm{BL}}\sin\theta| <$ 1.5 mm,
\item Minimum of six SCT clusters.
\end{itemize}

\subsection{Fit method}
\label{subsec:fit}

A measurement distribution of cluster \dEdx\ of tracks inside the jet core is fit using two \dEdx\ template distributions: a single-track template containing mainly tracks reconstructed from a single-particle cluster, and a multiple-track template mainly made up of tracks reconstructed from a merged cluster. Both templates are derived directly from collision data or from simulation for the corresponding efficiency measurements.

As verified in simulation, most highly collimated tracks are expected to be within $\deltaR<0.05$
which then defines the jet core for this method.
Outside the jet core, the contribution of collimated tracks is negligible, and therefore all tracks are expected to be reconstructed from a single-particle cluster. 
The single-track template is created using tracks reconstructed from clusters which are neither identified as merged nor shared and that are well outside the jet core (\deltaR$>0.1$). 
The multiple-track template is taken from tracks reconstructed from either B-layer clusters identified as merged or
shared B-layer clusters inside the jet core. These \MultUsed\ clusters are likely to be merged clusters.

Examples of the resulting distributions are shown in Figure~\ref{fig:dEdx_templates}. The single-track template, displayed as circles in Figure~\ref{fig:dEdx_templates}, contains a single peak at the \dEdx\ value expected for a MIP traversing the B-layer of the pixel detector and a long tail to higher values compatible with a Landau distribution. 
Contamination of merged clusters in this template is 0.3--0.5\% in the simulation.
The multiple-track template, displayed as squares in the same figure, instead exhibits a peak in the \dEdx\ range expected for two MIPs. A third, smaller peak occurs at \dEdx~\textgreater~3.2 \dEdxUnit for clusters created by three particles. The peak in the multiple-track template \dEdx\ distribution at values expected for one MIP is due to the fact that \MultUsed\ clusters can also originate from shared clusters or clusters identified as merged which, in truth, are not merged clusters.

\begin{figure}[h]
\centering
\includegraphics[width=0.49\textwidth]{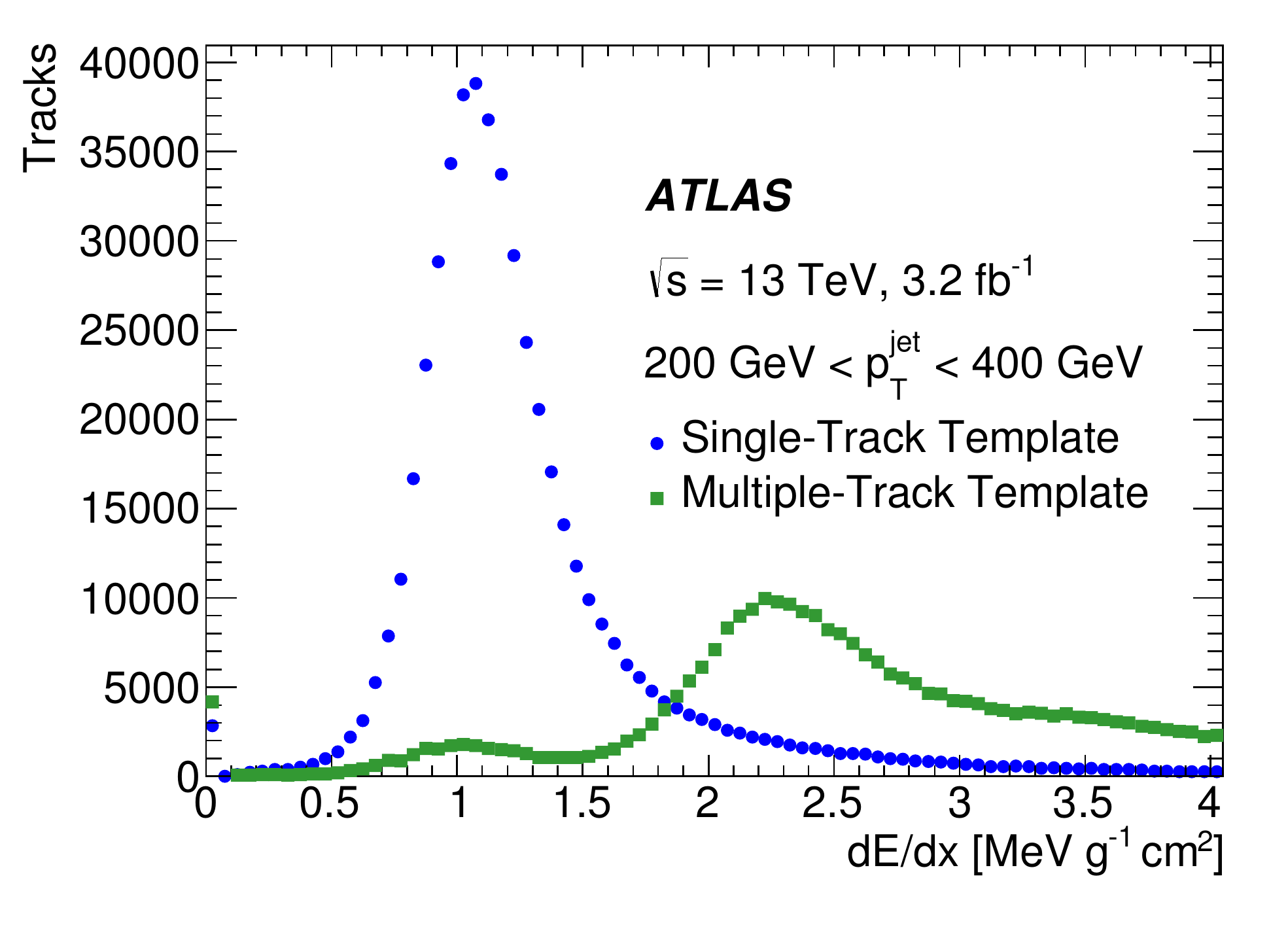}
\caption{Single-track and multiple-track templates for data with a jet \pT in the range $200~\GeV<$\ptJet\ $<400~\GeV$.}
\label{fig:dEdx_templates}
\end{figure}

The measurement distribution is created from tracks inside the jet core that are reconstructed from a cluster which is neither identified as merged nor shared. No additional requirements are made on other tracks using this cluster, including whether or not they satisfy the selections outlined in Section~\ref{subsec:selections}. The resulting \dEdx\ distribution contains single-particle clusters with a peak at the energy of one MIP and a long tail to high values, as well as an enhanced contribution of merged clusters from two particles. Contributions from clusters from more than two particles are negligible. The true two-MIP clusters are created from a pair of tracks where only one track is reconstructed. Therefore, for every reconstructed track in the measurement distribution with a merged cluster, there is one particle which is not reconstructed. 
Using this information, the number of tracks contributing to merged B-layer clusters from two particles ($N^{\textrm{True}}_{\textrm{2}}$) is found from the sum of the number of reconstructed particles in the multiple-track template ($N^{\textrm{Reco}}_{\textrm{2}}$) and twice the number of lost particles ($N_{\textrm{Lost}}$),
\begin{equation}
N^{\textrm{True}}_{\textrm{2}} = N^{\textrm{Reco}}_{\textrm{2}} + 2 \cdot N_{\textrm{Lost}}.
\end{equation}

The sample of $\rho$ decays discussed in Section~\ref{subsec:singleparticles} is used to confirm that the multiple-track template captures merged clusters and that the second MIP peak in the measurement sample does in fact contain merged clusters where one contributing particle is not reconstructed.
Therefore, to obtain the number of lost tracks ($N_{\textrm{Lost}}$), the measurement distribution is fit with the two templates. The fraction of merged clusters in the measurement distribution, $F^{\textrm{merged}}$, is
simply calculated from the post-fit number of tracks in the multiple-track template divided by the total number of tracks ($N^{\textrm{Reco}}_{\textrm{Data}}$). Finally, the fraction of lost tracks passing through the same detector element as a reconstructed track is given by:
\begin{eqnarray}
  \Fmerged &=& \frac{N_{\textrm{Lost}}}{N^{\textrm{True}}_{\textrm{2}}},  \\
        &\approx& \frac{N_{\textrm{Lost}}}{N^{\textrm{Reco}}_{\textrm{2}} + 2 \cdot N_{\textrm{Lost}}},
\label{eq:flost}
\end{eqnarray}
where
\begin{equation}
 N_{\textrm{Lost}}=F^{\textrm{merged}} \cdot N^{\textrm{Reco}}_{\textrm{Data}}.
\end{equation}
The relation is approximate due to the assumption that the lost track of a pair of tracks has the same properties (e.g. \pt and hit content) as the reconstructed track. In simulation, this assumption can be explicitly checked by requiring the truth particle corresponding to the lost track to also pass the analysis selections. This confirms that the deviation from the approximation results in a less than 1.5\% change in \Fmerged.

To minimize the effect of clusters created by more than two particles, the fit was performed over the range \numrange[range-phrase = --]{1.1}{3.07} (\numrange[range-phrase = --]{1.26}{3.2}) \dEdxUnit for data (simulation). Contributions from clusters from more than two particles in this range are of the order of a few percent. An offset in the distributions observed in MC events compared to data requires an adjustment of the respective fit ranges. The ranges are chosen to have the same fraction of clusters inside the fit range with respect to all clusters in the distribution. 
An imperfect description of the leading edge of the measurement distribution by the single-track template would affect the fitted result. Since the area of interest lies at much higher \dEdx\ values, the lower edge of the fit range was chosen to avoid as much as possible the leading edge of the single-particle \dEdx\ peak, while retaining a large sample for the remainder of the distribution. 

To study the dependence of lost tracks on jet \pT, the fit is performed in seven different bins of jet \pT ranging from 200~\GeV\ to 1600~\GeV\ in steps of 200~\GeV. 

The measurement is performed both on data and simulation samples. For simulation, separate templates are constructed for each jet-\pT bin. For data, the single-track and multiple-track templates are derived from the lowest jet-\pT bin, shown in Figure~\ref{fig:dEdx_templates}, due to the small number of events at higher jet \pT. It was verified that within the statistical uncertainty of the high-\pT bins, the templates derived from the lowest jet-\pT bin have the same shape within the fitted range. 

\subsection{Systematic uncertainties}
\label{subsec:systematics}

The resulting $F^{\textrm{merged}}$ exhibits a statistical uncertainty due to the finite number of entries in both the template and the measurement distributions. 

Various potential sources of systematic bias were studied and are discussed below. The relative values for data are summarized in Table~\ref{tab:relSysValues} and values for MC simulation are comparable.
The measured \Fmerged\ varies as a function of the range in \dEdx\ for which the distribution is fit. This is due to the different fractions of clusters with a \dEdx\ of two and three MIPs falling in the fitted range. The effect was estimated by increasing the fit range. The fitting process was repeated for six different ranges with the upper edge increasing in 0.2 \dEdxUnit\ increments. A symmetric uncertainty, equal to the maximum change in \Fmerged, is applied to each jet-\pT bin. The start of the fitted range was chosen such that small variations have a negligible impact on \Fmerged.

A systematic uncertainty considered for data is the result of fitting all data jet-\pT bins with the templates from the lowest jet-\pT bin. This results in an overestimate of \Fmerged\ increasing with jet \pT.
To account for this bias, a \pT-dependent multiplicative correction was determined by comparing the \Fmerged\ values fitted in simulation with templates from the corresponding jet-\pT bin
with those obtained using a template from the lowest jet-\pT bin. This correction increases from about 10\% to 25\% for jets with a \pT ranging from 400~\GeV\ to 600~\GeV\ and from 1400~\GeV\ to 1600~\GeV, respectively.  
This correction term was applied to data \Fmerged\ values after completing the fitting procedure. In addition, the difference between the two simulation \Fmerged\ values compared for the correction factor was also included as a systematic uncertainty.
An additional check performed with a large simulated sample showed a 3\% to 8\% bias in \Fmerged\ in the studied jet-\pT\ range due to the fraction of tracks reconstructed from $\geq$ 3 particle clusters, relative to the two-particle contribution in the multiple-track template. 

To validate the method, and provide an estimate of any residual biases, 
a truth-based closure test was performed using simulated samples. 
At low jet \pt, the residual \dEdx\ peak at values expected from one MIP 
in the multiple-track template contributes to a non-closure. 
Also, for all jet $p_\textrm{T}$,
isolated-track reconstruction efficiency, the composition of multiple-particle clusters, including particle composition 
and the calibration of \dEdx\ itself are all covered
in this non-closure estimate.
This is already covered by the systematic uncertainty determined from changing the fit range
described above, but also leads to a non-closure.
In the lowest jet-\pT bin, a non-closure of approximately +18\% is observed,
corresponding to an absolute overestimation of the true \Fmerged\ of about 0.013,
but then quickly decreases with increasing jet \pT.
This uncertainty is included for both simulation and data with
the corresponding relative values in Table~\ref{tab:relSysValues}. 

Other possible sources of uncertainty are contributions to \Fmerged\ not originating from the density of the environment.
Such contributions could come from pile-up tracks creating merged clusters with tracks in the jets, as well as lost isolated tracks.
Conservative estimates based on MC studies showed that such contributions are 2\% to 6\% of the total \Fmerged\ in the studied jet-\pT range. This effect is covered by the non-closure systematic uncertainty described above.

Uncertainties in the jet energy scale calibration and resolution have negligible impact in the analysis. Possible effects due to the binning of the \dEdx\ distributions were studied and found also to be insignificant.

\begin{table}
\begin{center}
\begin{tabular}{r|cccccc}
\hline \hline
 Jet \pT [\GeV]       & \Fmerged & Fit range [\%] & Low-\pT temp. [\%] & Non-closure [\%] & Tot. syst. [\%] & Stat. [\%] \\
\hline
200--400   & 0.061 & 13 &  0 & 18 & 23 & 10 \\
400--600   & 0.063 & 12 &  7 & 11 & 17 & 6 \\
600--800   & 0.070 & 10 & 13 & 6  & 17  & 7 \\
800--1000  & 0.064 & 12 & 18 & 1  & 22  & 11 \\
1000--1200 & 0.067 & 12 & 21 & 0  & 24  & 15 \\
1200--1400 & 0.080 & 11 & 16 & 0  & 19  & 13 \\
1400--1600 & 0.093 & 15 & 16 & 0  & 22  & 18 \\
\hline \hline
\end{tabular}
\end{center}
\caption{Measured \Fmerged, relative values of leading systematic uncertainties, and total systematic and statistical uncertainty in the fraction of lost tracks for data in bins of jet \pT.}
\label{tab:relSysValues}
\end{table}

\subsection{Results}
\label{subsec:results}

Figure~\ref{fig:dEdx_dataFits} shows the fit result for data in two bins of jet \pt. The single-track and multiple-track \dEdx\ templates provide a good description of the \dEdx\ distribution as visible from the ratio in Figure~\ref{fig:dEdx_dataFits}.

\begin{figure}[h]
\begin{center}
\subfigure[]{\includegraphics[width=0.49\textwidth]{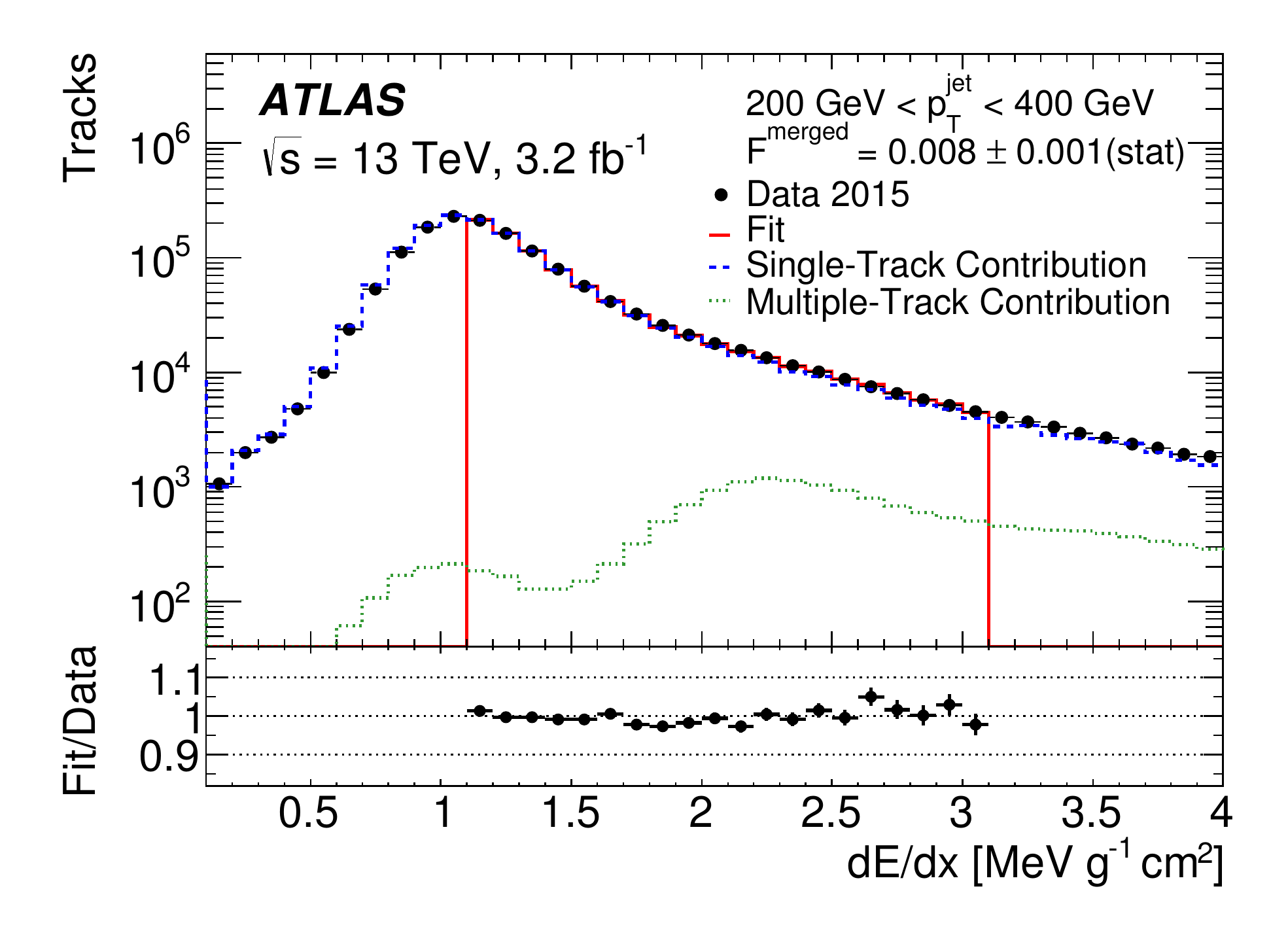}}
\subfigure[]{\includegraphics[width=0.49\textwidth]{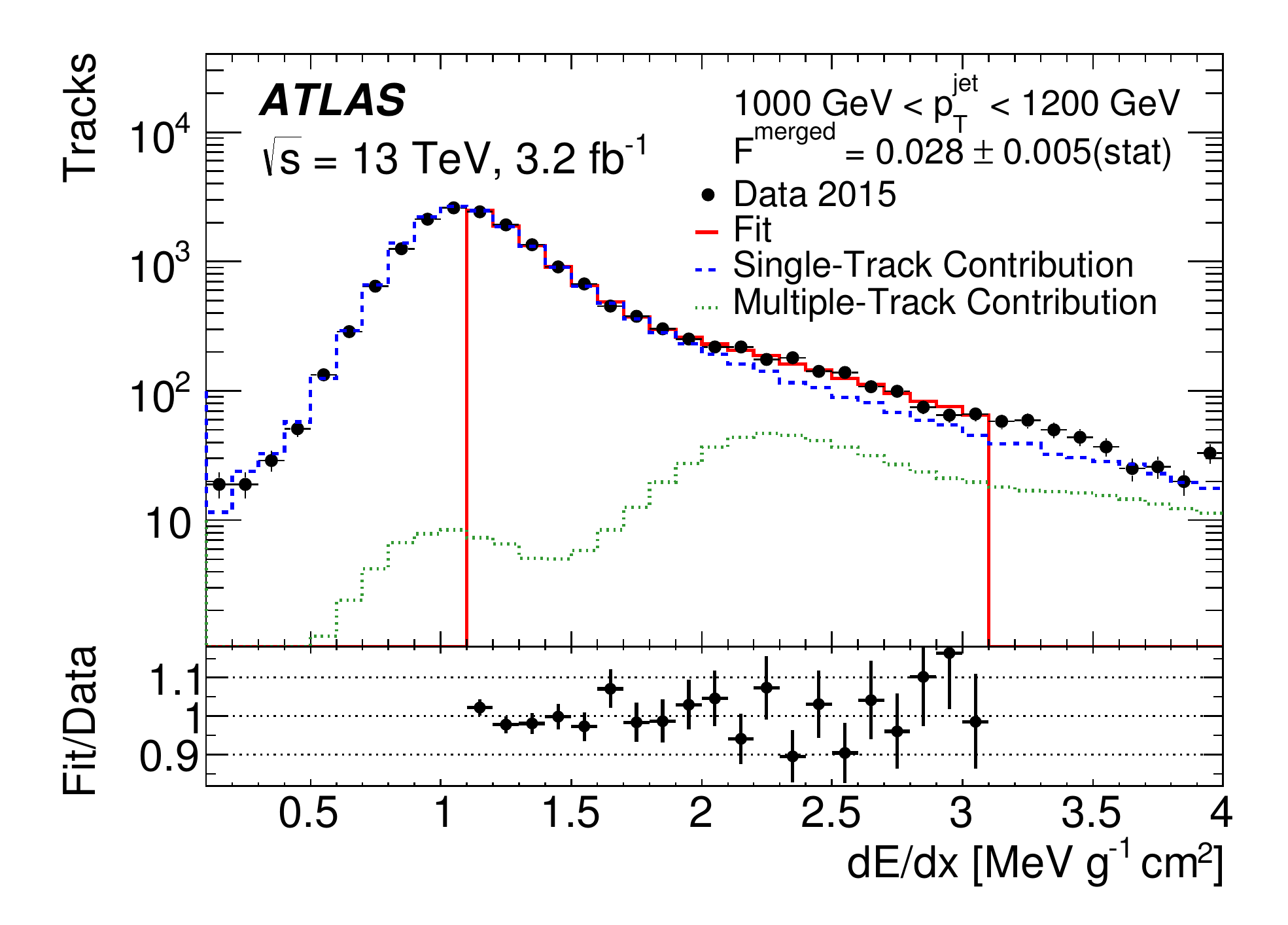}}
\end{center}
\caption{Data \dEdx\ measurement distributions (black circles) with fit results (solid line) are shown for (a) $200~\GeV<\pt^{\mathrm{jet}}<400~\GeV$ and (b) $1000~\GeV<\pt^{\mathrm{jet}}<1200~\GeV$.  The single-track template scaled by $1 - F^{\textrm{merged}}$ is shown as the single-track contribution (dashed line) and the multiple-track template scaled by $F^{\textrm{merged}}$ is shown as the multiple-track contribution (dotted line). The bottom panel in each plot shows the ratio of the fit to the data within the fit range (1.1--3.07 \dEdxUnit). }
\label{fig:dEdx_dataFits}
\end{figure}

Differences between event generators, such as different hadronization models and flavour compositions, can affect \Fmerged\ and the overall comparison of data and MC simulation.
By comparing the fit results from simulated samples made with the \PYTHIA 8, \SHERPA and \Herwigpp event generators, a generator uncertainty was derived for simulation only. For each jet-\pT bin, results from \PYTHIA are taken as the central value and the largest difference of \Fmerged\ between the three generators is symmetrized and taken as the generator uncertainty. The relative generator uncertainties in the fraction of lost tracks ranges from 4\% to 37\% in the different jet-\pT bins.

A comparison of \Fmerged\ as a function of jet \pt for data and simulation is shown in Figure~\ref{fig:dEdx_Fmerged}. As the jet \pt increases, so does \Fmerged, with a similar trend observed in both data and simulation. This increase is caused by an increasing density of charged particles, which thereby causes higher collimation of the track pair, and is not due to confusion in correctly assigning clusters to tracks. At a certain point, the two particles are so collimated that the reconstructed tracks start to overlap completely up to the radius of the SCT detector. At that point a similar effect as shown for tracks from the $\rho$ decay in Figures~\ref{fig:hitEffeta} and \ref{fig:tauRecEffNoShared} occurs. The cluster assignment efficiency for reconstructed tracks remains constant with increasing jet \pt, indicating no degradation of performance due to the environmental effects besides the second track. Only because of their increasingly collimated nature, the probability of losing one of the tracks rises. This effect was confirmed in simulation for tracks selected by this analysis.

The measurements in data and MC simulation are consistent across the whole studied jet-\pt range.

\begin{figure}[h]
\begin{center}
\includegraphics[width=0.49\textwidth]{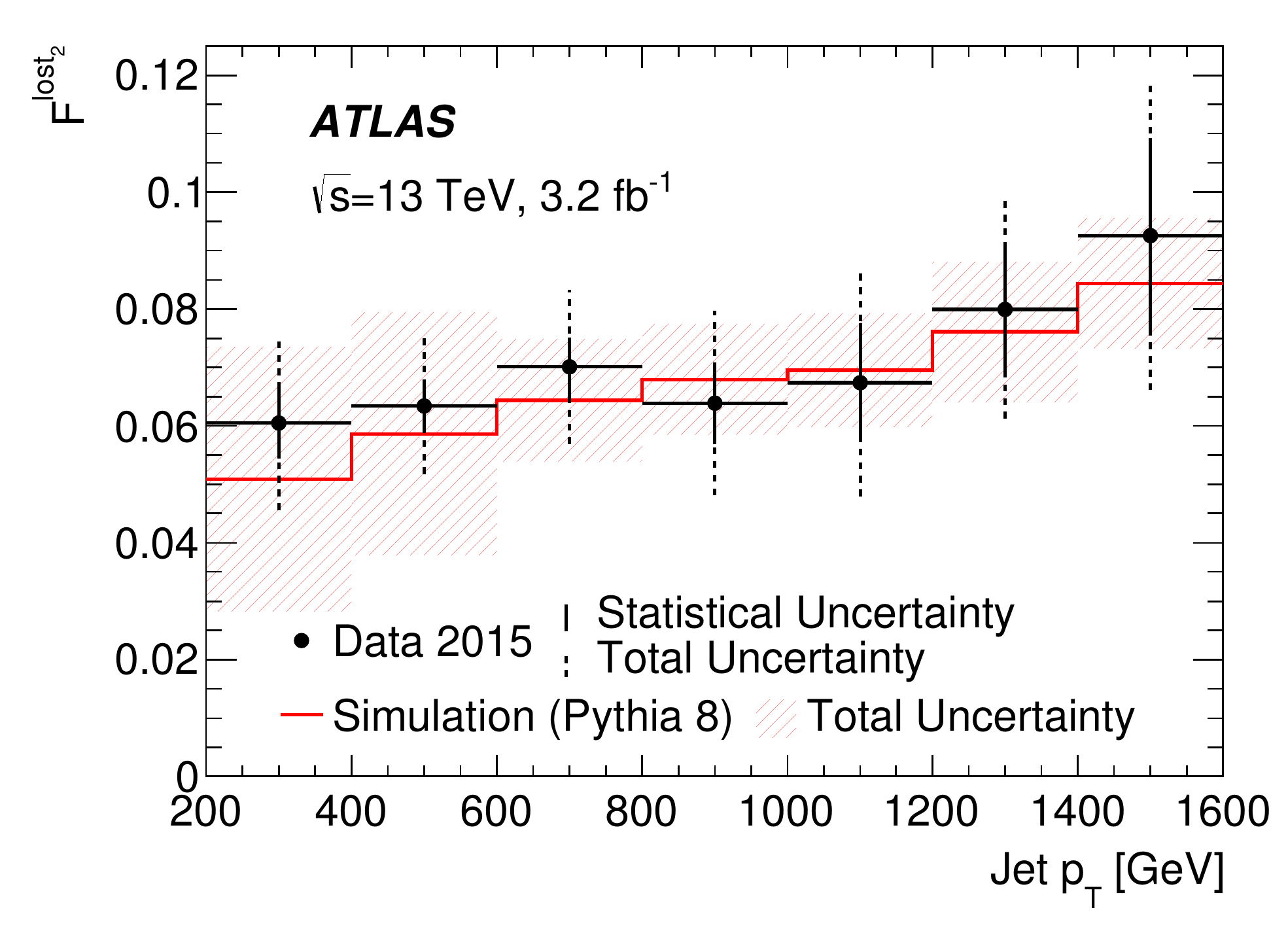}
\end{center}
\caption{The measured fraction of lost tracks, \Fmerged, in the jet core (\deltaR$<0.05$) as a function of jet \pt for data (black circles) and simulation (red line). Vertical solid error bars indicate statistical uncertainty, while the total uncertainty is represented by dashed error bars for data and a shaded area for simulation.}
\label{fig:dEdx_Fmerged}
\end{figure}

\FloatBarrier
\section{Conclusion}
\label{sec:conclusion}
This paper presents the performance of the ATLAS track reconstruction chain with
detailed studies in dedicated topologies, such as the cores of high-\pt\ 
jets and the decays of $\tau$-leptons, that are characterized by charged-particle separations comparable to
the inner detector's sensor granularity. The ambiguity-solver stage of the reconstruction
chain is described, including the usage of a neural-network-based approach
to identify pixel clusters created by multiple charged particles.
The current performance is demonstrated with simulated samples of a single particle decaying to a set of collimated charged particles.
In the cores of jets, the number of IBL clusters on tracks, as well as the expected track reconstruction efficiency, is robust up to the highest investigated \pt values.

A novel, fully data-driven technique, using the energy loss to identify clusters as originating from two charged particles is introduced to measure the fraction of charged 
particles, creating these clusters, that fail to be reconstructed.
The results are presented using tracks with \pT\ above 10~\GeV\ in the core of 
a jet from 3.2~fb$^{-1}$ of 13~\TeV\ proton--proton collisions at the LHC.
The measured fraction of lost tracks as a function of jet transverse momentum was found to range from $0.061 \pm 0.006\textrm{(stat.)} \pm 0.014\textrm{(syst.)}$ to $0.093 \pm 0.017\textrm{(stat.)}\pm 0.021\textrm{(syst.)}$ as the jet \pt increases from 200~\GeV\ to 1600~\GeV. Data and simulation are compatible for the full studied jet-\pt range. This result can be used to minimize the uncertainty in the track reconstruction inefficiency in the cores of jets relevant for jet energy and mass calibrations as well as
measurements of jet properties.

\section*{Acknowledgements}

We thank CERN for the very successful operation of the LHC, as well as the
support staff from our institutions without whom ATLAS could not be
operated efficiently.

We acknowledge the support of ANPCyT, Argentina; YerPhI, Armenia; ARC, Australia; BMWFW and FWF, Austria; ANAS, Azerbaijan; SSTC, Belarus; CNPq and FAPESP, Brazil; NSERC, NRC and CFI, Canada; CERN; CONICYT, Chile; CAS, MOST and NSFC, China; COLCIENCIAS, Colombia; MSMT CR, MPO CR and VSC CR, Czech Republic; DNRF and DNSRC, Denmark; IN2P3-CNRS, CEA-DSM/IRFU, France; SRNSF, Georgia; BMBF, HGF, and MPG, Germany; GSRT, Greece; RGC, Hong Kong SAR, China; ISF, I-CORE and Benoziyo Center, Israel; INFN, Italy; MEXT and JSPS, Japan; CNRST, Morocco; NWO, Netherlands; RCN, Norway; MNiSW and NCN, Poland; FCT, Portugal; MNE/IFA, Romania; MES of Russia and NRC KI, Russian Federation; JINR; MESTD, Serbia; MSSR, Slovakia; ARRS and MIZ\v{S}, Slovenia; DST/NRF, South Africa; MINECO, Spain; SRC and Wallenberg Foundation, Sweden; SERI, SNSF and Cantons of Bern and Geneva, Switzerland; MOST, Taiwan; TAEK, Turkey; STFC, United Kingdom; DOE and NSF, United States of America. In addition, individual groups and members have received support from BCKDF, the Canada Council, CANARIE, CRC, Compute Canada, FQRNT, and the Ontario Innovation Trust, Canada; EPLANET, ERC, ERDF, FP7, Horizon 2020 and Marie Sk{\l}odowska-Curie Actions, European Union; Investissements d'Avenir Labex and Idex, ANR, R{\'e}gion Auvergne and Fondation Partager le Savoir, France; DFG and AvH Foundation, Germany; Herakleitos, Thales and Aristeia programmes co-financed by EU-ESF and the Greek NSRF; BSF, GIF and Minerva, Israel; BRF, Norway; CERCA Programme Generalitat de Catalunya, Generalitat Valenciana, Spain; the Royal Society and Leverhulme Trust, United Kingdom.

The crucial computing support from all WLCG partners is acknowledged gratefully, in particular from CERN, the ATLAS Tier-1 facilities at TRIUMF (Canada), NDGF (Denmark, Norway, Sweden), CC-IN2P3 (France), KIT/GridKA (Germany), INFN-CNAF (Italy), NL-T1 (Netherlands), PIC (Spain), ASGC (Taiwan), RAL (UK) and BNL (USA), the Tier-2 facilities worldwide and large non-WLCG resource providers. Major contributors of computing resources are listed in Ref.~\cite{ATL-GEN-PUB-2016-002}.

\clearpage
\printbibliography

\newpage
\begin{flushleft}
{\Large The ATLAS Collaboration}

\bigskip

M.~Aaboud$^\textrm{\scriptsize 137d}$,
G.~Aad$^\textrm{\scriptsize 88}$,
B.~Abbott$^\textrm{\scriptsize 115}$,
J.~Abdallah$^\textrm{\scriptsize 8}$,
O.~Abdinov$^\textrm{\scriptsize 12}$$^{,*}$,
B.~Abeloos$^\textrm{\scriptsize 119}$,
S.H.~Abidi$^\textrm{\scriptsize 161}$,
O.S.~AbouZeid$^\textrm{\scriptsize 139}$,
N.L.~Abraham$^\textrm{\scriptsize 151}$,
H.~Abramowicz$^\textrm{\scriptsize 155}$
% ...
\bigskip
\\
% ...
$^{8}$ Department of Physics, The University of Texas at Arlington, Arlington TX, United States of America\\
% ...
$^{12}$ Institute of Physics, Azerbaijan Academy of Sciences, Baku, Azerbaijan\\
% ...
$^{88}$ CPPM, Aix-Marseille Universit{\'e} and CNRS/IN2P3, Marseille, France\\
% ...
$^{115}$ Homer L. Dodge Department of Physics and Astronomy, University of Oklahoma, Norman OK, United States of America\\
% ...
$^{119}$ LAL, Univ. Paris-Sud, CNRS/IN2P3, Universit{\'e} Paris-Saclay, Orsay, France\\
% ...
$^{137}$ $^{(a)}$ Facult{\'e} des Sciences Ain Chock, R{\'e}seau Universitaire de Physique des Hautes Energies - Universit{\'e} Hassan II, Casablanca; $^{(b)}$ Centre National de l'Energie des Sciences Techniques Nucleaires, Rabat; $^{(c)}$ Facult{\'e} des Sciences Semlalia, Universit{\'e} Cadi Ayyad, LPHEA-Marrakech; $^{(d)}$ Facult{\'e} des Sciences, Universit{\'e} Mohamed Premier and LPTPM, Oujda; $^{(e)}$ Facult{\'e} des sciences, Universit{\'e} Mohammed V, Rabat, Morocco\\
% ...
$^{139}$ Santa Cruz Institute for Particle Physics, University of California Santa Cruz, Santa Cruz CA, United States of America\\
% ...
$^{151}$ Department of Physics and Astronomy, University of Sussex, Brighton, United Kingdom\\
% ...
$^{155}$ Raymond and Beverly Sackler School of Physics and Astronomy, Tel Aviv University, Tel Aviv, Israel\\
% ...
$^{161}$ Department of Physics, University of Toronto, Toronto ON, Canada\\
% ...
$^{a}$ Also at Department of Physics, King's College London, London, United Kingdom\\
$^{b}$ Also at Institute of Physics, Azerbaijan Academy of Sciences, Baku, Azerbaijan\\
$^{c}$ Also at Novosibirsk State University, Novosibirsk, Russia\\
$^{d}$ Also at TRIUMF, Vancouver BC, Canada\\
$^{e}$ Also at Department of Physics {\&} Astronomy, University of Louisville, Louisville, KY, United States of America\\
% ...
$^{*}$ Deceased
\end{flushleft}



\end{document}